\documentclass[prx,twocolumn,showpacs,amsmath,amssymb,superscriptaddress,floatfix,reprint, nofootinbib]{revtex4-2}
\usepackage{amsmath}
\usepackage{amssymb}
\usepackage{amsthm}
\usepackage{amsfonts}
\usepackage{dsfont}
\usepackage{listings}
\usepackage{macros}
\usepackage{enumerate}
\usepackage{latexsym}
\usepackage{tocenter}
\usepackage{hyperref}

\usepackage{bbold}

\usepackage{psfrag}
\usepackage{xcolor}
\usepackage{float}

\usepackage{comment}

\usepackage{graphicx}
\usepackage{subfigure}

\makeatletter
\def\l@subsubsection#1#2{}
\makeatother

\begin{document}

\tolerance 10000
\title{Numerical Methods for Detecting Symmetries and Commutant Algebras}
\author{Sanjay Moudgalya}
\email{sanjaym@caltech.edu}
\affiliation{Department of Physics and Institute for Quantum Information and Matter,
California Institute of Technology, Pasadena, California 91125, USA}
\affiliation{Walter Burke Institute for Theoretical Physics, California Institute of Technology, Pasadena, California 91125, USA}
\author{Olexei I. Motrunich}
\affiliation{Department of Physics and Institute for Quantum Information and Matter,
California Institute of Technology, Pasadena, California 91125, USA}
\begin{abstract}
For \textit{families} of Hamiltonians defined by parts that are local, the most general definition of a symmetry algebra is the commutant algebra, i.e., the algebra of operators that commute with each local part.
Thinking about symmetry algebras as commutant algebras allows for the treatment of conventional symmetries and unconventional symmetries (e.g., those responsible for weak ergodicity breaking phenomena) on equal algebraic footing.
In this work, we discuss two methods for numerically constructing this commutant algebra starting from a family of Hamiltonians.
First, we use the equivalence of this problem to that of simultaneous block-diagonalization of a given set of local operators, and discuss a probabilistic method that has been found to work with probability $1$ for both Abelian and non-Abelian symmetries or commutant algebras. 
Second, we map this problem onto the problem of determining frustration-free ground states of certain Hamiltonians, and we use ideas from tensor network algorithms to efficiently solve this problem in one dimension.
These numerical methods are useful in detecting standard and non-standard conserved quantities in families of Hamiltonians, which includes examples of regular symmetries, Hilbert space fragmentation, and quantum many-body scars, and we show many such examples.
In addition, they are necessary for verifying several conjectures on the structure of the commutant algebras in these cases, which we have put forward in earlier works~\cite{moudgalya2021hilbert, moudgalya2022from, moudgalya2022exhaustive}.
Finally, we also discuss similar methods for the inverse problem of determining local operators with a given symmetry or commutant algebra, which connects to existing methods in the literature.
A special case of this construction reduces to well-known ``Eigenstate to Hamiltonian" methods for constructing Hermitian local operators that have a given state as an eigenstate.
\end{abstract}
\date{\today}
\maketitle
%

\tableofcontents

\section{Introduction}
\label{sec:intro}
The study of symmetries is at the heart of many areas of physics. 
For example, understanding of symmetries leads to classifications of various phases of matter and transitions between them~\cite{sachdev2011quantum, fradkin2013field, zeng2019quantum}, definitions of Gibbs ensembles that describe thermodynamics of systems~\cite{d2016quantum, mori2018thermalization}, etc.
Hence, given the Hamiltonian for a quantum system, determining its symmetries is of utmost importance and is usually the first step in approaching any problem.  
Most symmetries studied in quantum many-body physics, such as $Z_2$ parity symmetry, $U(1)$ particle number conservation, or $SU(2)$ spin conservation, are examples of ``on-site" symmetries represented by global unitary operators that are tensor products of single-site unitary operators that form representations of the corresponding groups; and symmetries of this form are usually easy to ``guess" by quick direct inspection of the Hamiltonian.
However, a relatively recent realization is that not all symmetries that appear in natural Hamiltonians are of this type, and the exploration beyond such ``conventional" on-site global symmetries has only recently been initiated in various contexts~\cite{aasen2016topological, aasen2020topological, quella2020symmetry,lootens2021MPO,  moudgalya2021hilbert, lootens2021dualities, chatterjee2022algebra, mcgreevy2022generalized, moudgalya2022from, moudgalya2022exhaustive}.
Particularly in the context of dynamics, invoking such unconventional symmetries has also been shown to be necessary to understand phenomena of weak ergodicity breaking~\cite{serbyn2020review, papic2021review, moudgalya2021review, sengupta2021phases, chandran2022review} that have now been observed in multiple experiments~\cite{bernien2017probing, scherg2020observing, bluvstein2021controlling, kohlert2021experimental, su2022observation}.
The study of unconventional symmetries first requires a clear general definition of a symmetry, which is not immediately obvious.
Allowing arbitrary operators that commute with the given Hamiltonian to be valid conserved quantities implies that any Hamiltonian has exponentially many conserved quantities, namely, its eigenstate projectors are all trivially conserved. 
In \cite{moudgalya2021hilbert, moudgalya2022from, moudgalya2022exhaustive} we introduced and developed one framework that provides a non-trivial definition of symmetries and conserved quantities, which naturally captures both conventional symmetries in standard Hamiltonians~\cite{moudgalya2022from} as well as unconventional symmetries that explain the weak ergodicity breaking phenomena of quantum many-body scars~\cite{moudgalya2022exhaustive} and Hilbert space fragmentation~\cite{moudgalya2021hilbert}. 
In particular, we focus on \textit{families} of Hamiltonians that are composed of local terms, and the bond or local algebra $\mA$ generated by the local terms~\cite{nussinov2009bond, moudgalya2021hilbert, moudgalya2022from, moudgalya2022exhaustive}.
The commutant algebra $\mC$ is the centralizer of the local algebra, i.e.,  the algebra of all conserved quantities of the said family of Hamiltonians, which we will also refer to as the symmetry algebra. 
Hence we can think of any symmetry as being characterized by a pair of algebras $(\mA, \mC)$, which are simply the algebras of symmetric operators and the symmetry algebra respectively. 
This definition of conserved quantities starting from an arbitrary local algebra $\mA$ removes the restriction to on-site unitary and other conventional symmetries, since the commutant $\mC$ can include non-on-site or even non-local operators, and moreover it need not have any simple underlying group structure, as illustrated through several examples in \cite{moudgalya2021hilbert, moudgalya2022exhaustive}.
In addition, this language allows a systematic counting of the number and the sizes of dynamically disconnected blocks or Krylov subspaces in terms of the dimensions of the commutant, or more precisely the dimensions of the irreducible representations of the local algebra and the commutant.
This in turn leads to a broad classification of systems in terms of the scaling of the number of Krylov subspaces with system size~\cite{moudgalya2021hilbert} (or equivalently their Coherence Generating Power~\cite{zanardi2017coherence, andreadakis2022coherence}), as well as to a precise definition of Hilbert space fragmentation~\cite{moudgalya2021hilbert}.
Moreover, a complete determination of the \textit{full} commutant $\mC$ corresponding to a local algebra $\mA$ enables an exhaustive characterization of all symmetric local operators.
That is, the von Neumann Double Commutant Theorem (DCT) guarantees that \textit{any} operator that commutes with all operators in $\mC$ should be in the local algebra $\mA$, hence it should be expressible in terms of the known local generators of $\mA$, which enables systematic construction and characterization of \textit{all} local operators with a given symmetry.
We demonstrated this principle in action by constraining the forms of operators symmetric under several regular symmetries~\cite{moudgalya2022from}, and also by  constructing local Hamiltonians that possess some set of Quantum Many-Body Scars~\cite{moudgalya2022exhaustive}.
However, unlike simple on-site symmetries, determining the full commutant $\mC$ corresponding to a local algebra $\mA$ or determining the dimensions of their irreducible representations is far from straightforward in practice.
Moreover, we need these properties for many of the results we discuss in previous works, e.g., the Hilbert space decomposition and the application of the DCT require an explicit proof that a particular set of operators generates the \textit{full} commutant, and not just some subalgebra of the true commutant. 
As evident from examples in \cite{moudgalya2022from, moudgalya2022exhaustive}, we were able to analytically establish the commutants only in some simple cases, and in many cases we also needed to invoke some reasonable assumptions. 
Hence in this work, we discuss some numerical methods necessary to construct the commutant $\mC$ corresponding to a local algebra $\mA$, which can either be used to determine and numerically construct operators in $\mC$, or to quickly verify key properties, e.g., the dimension of the $\mC$.
One approach we discuss is to view this as problem of simultaneous block diagonalization of the (non-commuting) generators of the local algebra $\mA$.
The problem of simultaneous block diagonalization appears in several other contexts~\cite{holbrook2003noiseless, choi2006method, wang2013numerical, chertkov2020engineering, nie2021operator, bondarenko2021constructing, murota2010numerical, maehara2010numerical}, notably in the study of ``noise commutants" and ``noiseless subsystems" in the quantum error correction literature~\cite{holbrook2003noiseless, choi2006method, wang2013numerical}.
However, some of these methods are not always directly applicable or completely general, e.g., the methods proposed to detect symmetries in \cite{chertkov2020engineering, nie2021operator} only work for Abelian symmetries, hence we build upon them and adapt them to the quantum many-body problem we have in hand, including with non-Abelian symmetries.
In another approach, we also exploit the \textit{locality} of the generators of $\mA$ and map the problem of finding the commutant to determining the ground state of a frustration-free Hamiltonian, which is a superoperator acting on the space of all operators.
This method is applicable is any dimension, but the brute force implementation is computationally more expensive than the previous method, since it required the diagonalization of a superoperator.
However, for one-dimensional systems, this mapping enables us to use ideas from Matrix Product State (MPS) algorithms to obtain an efficient method to determine the commutant $\mC$, building on methods previously used for related problems~\cite{debeaudrap2010solving, movassagh2010unfrustrated, yao2022bounding}. 
Finally, for the sake of completeness, we also discuss the ``inverse" of our method, which constructs local symmetric operators corresponding to a given commutant algebra.
Remarkably, this coincides with and also unifies many of the existing methods used in the literature for constructing local operators with a given symmetry~\cite{obrien2016explicit, chertkov2020engineering} or those that possess a particular eigenstate~\cite{qi2019determininglocal, chertkov2018computational, yang2022detecting, greiter2018method, dalmonte2018quantum, turkeshi2019entanglement, pakrouski2020automatic, bairey2019learning, turkeshi2020parent}.
Since such local operators are in the algebra $\mA$ corresponding to a given commutant $\mC$, such methods are also useful in determining a set of local generators of $\mA$.
In addition, these methods are useful in systematically constructing two distinct types of symmetric operators,  which we referred to as  type I and type II symmetric operators in an earlier work~\cite{moudgalya2022exhaustive},  and we discuss such applications.
This paper is organized as follows.
In Sec.~\ref{sec:commutants}, we briefly review the concepts of bond, local, and commutant algebras and Hilbert space decomposition that we use in the rest of this work.
In Sec.~\ref{sec:commutantblock}, we discuss a method for simultaneous block diagonalization of operators in the local algebra and using it to construct operators in the commutant.
In Sec.~\ref{sec:commutantLiouv}, we discuss a method for constructing the commutant as a tensor network, and we illustrate efficient method to obtain the commutant in one dimension.
Finally, in Sec.~\ref{sec:commutantloc}, we discuss numerical methods that construct local operators in the local algebra given its commutant, and we conclude with open questions in Sec.~\ref{sec:conclusions}.
\section{Bond, Local, and Commutant Algebras}\label{sec:commutants}
We now briefly review the concepts of local and commutant algebras required for this work, and we refer to our previous works~\cite{moudgalya2021hilbert, moudgalya2022from, moudgalya2022exhaustive} for a detailed discussion of their properties.
Given a set of generically non-commuting operators $\{\hH_\alpha\}$ on a Hilbert space $\mH$, we can define its \textit{commutant} to be the set of all operators $\hO$ that satisfy
\begin{equation}
    [\hH_\alpha, \hO] = 0\;\;\;\forall \alpha.     
\label{eq:eachtermcommute}
\end{equation}
These operators $\hO$ form an associative algebra $\mC$, which we refer to as the commutant algebra, which can also be viewed as the centralizer of the algebra generated by the operators $\{\hH_\alpha\}$, denoted by $\mA \defn \lgen \{\hH_\alpha\} \rgen$, where the notation implicitly assumes inclusion of the identity operator $\mathds{1}$ and also closure under Hermitian conjugation. 
$\mA$ and $\mC$ are hence examples of von Neumann algebras~\cite{landsman1998lecture, harlow2017, kabernik2021reductions}, and are centralizers of each other as a consequence of the DCT.
As a consequence, the centers of the algebras $\mA$ and $\mC$ (i.e., the subalgebra that commutes with all the elements in the algebra) coincide, and can be written as $\mZ = \mA \cap \mC$; and when $\mC$ or $\mA$ is Abelian, we obtain that $\mZ = \mC \subseteq \mA$ or $\mZ = \mA \subseteq \mC$ respectively.
In quantum matter applications, we are interested in cases where the operators $\{\hH_\alpha\}$ are local, which can refer to either strictly local operators with a support over a few nearby sites, or extensive local operators that are sums of strictly local terms throughout the system.
Consequently, we refer to the algebra $\mA$ generated by these terms as a bond algebra if $\{\hH_\alpha\}$ contains only strictly local terms, or more generally as a local algebra if it contains some extensive local terms.  
The commutant algebra $\mC$ is then naturally the symmetry algebra of \textit{families of} Hamiltonians~\cite{moudgalya2021hilbert, moudgalya2022from} of the form
\begin{equation}
    H = \sumal{\alpha}{}{J_{\alpha} \hH_{\alpha}}, 
\label{eq:genhamil}
\end{equation}
where $\{J_\alpha\}$ is an arbitrary set of coefficients.
As we discuss in other works, the language of commutant algebras can be used to understand a variety of phenomena such as Hilbert space fragmentation~\cite{moudgalya2021hilbert}, conventional symmetries~\cite{moudgalya2022from}, and quantum many-body scars~\cite{moudgalya2022exhaustive}. 
In all of these cases, $\mC$ and $\mA$ are respectively the ``symmetry" algebra (i.e., all conserved quantities) and the algebra of ``symmetric" operators (i.e., all operators that commute with the conserved quantities in $\mC$).
Note that the commutant $\mC$ should be interpreted as the symmetry algebra for \textit{generic} families of Hamiltonians constructed from the set $\{\hH_\alpha\}$.
We usually start with a set $\{\hH_\alpha\}$ that is ``homogeneously" distributed over the lattice, which is physically reasonable and captures physically relevant symmetries.
Breaking homogeneity and excluding some operators from the set might lead to a larger commutant, which, while mathematically correct, is not physically relevant. 
For example, in some cases this exclusion might effectively ``split" the system into some commuting parts, giving rise to several ``unwanted" symmetries.
However, with judicious choices, it is easy to avoid such situations.
An important property that we heavily use in this work is the decomposition of the Hilbert space as a consequence of the existence of algebras $\mA$ and $\mC$ that are centralizers of each other~\cite{zanardi2001virtual, bartlett2007reference, lidar2014dfs, moudgalya2021hilbert, moudgalya2022from, moudgalya2022exhaustive, fulton2013representation}, given by
\begin{equation}
    \mH = \bigoplus_{\lambda}{\left(\mH^{(\mA)}_\lambda \otimes \mH^{(\mC)}_\lambda\right)},
\label{eq:Hilbertdecomp}
\end{equation}
where $\mH^{(\mA)}_\lambda$ and $\mH^{(\mC)}_\lambda$ respectively denote $D_\lambda$- and $d_\lambda$-dimensional irreps of $\mA$ and $\mC$.
Equation~(\ref{eq:Hilbertdecomp}) can be simply viewed as a tensored basis in which all the operators in $\mA$ are simultaneously (maximally) block-diagonal.
Operationally this means that any operators $\hh_\mA \in \mA$, $\hh_{\mC}\in \mC$, and $\hh_{\mZ} \in \mZ$ can be unitarily transformed as
\begin{align}
    &W^\dagger \hh_{\mA} W = \bigoplus_{\lambda} (M^\lambda(\hh_\mA) \otimes \mathds{1}_{d_\lambda}),\label{eq:Ablock}\\
    &W^\dagger \hh_{\mC} W = \bigoplus_{\lambda} (\mathds{1}_{D_\lambda} \otimes N^\lambda(\hh_\mC)), \label{eq:Cblock} \\
    &W^\dagger \hh_{\mZ} W = \bigoplus_\lambda(c_\lambda(\hh_{\mZ}) \mathds{1}_{D_\lambda} \otimes \mathds{1}_{d_\lambda}),
\label{eq:Zblock}
\end{align}
where $M^\lambda(\hh_\mA)$ and $N^\lambda(\hh_\mC)$ are some $D_\lambda$-dimensional and $d_\lambda$-dimensional matrices respectively, $c_\lambda(\hh_{\mZ})$ is a c-number, and $W$ is a fixed unitary matrix.
Equation~(\ref{eq:Ablock}) precisely denotes the \textit{simultaneous} block-diagonalization of the operators in (including the generators of) $\mA$, which is equivalent to the simultaneously block-diagonalization of operators in (including the generators of) $\mC$, shown in Eq.~(\ref{eq:Cblock}). 
Since the Hamiltonians we are interested in belong to the local algebra $\mA$, this decomposition can be used to precisely define dynamically disconnected quantum number sectors or ``Krylov subspaces" of the Hamiltonian~\cite{moudgalya2021hilbert}.
In particular, there are blocks labelled by different $\lambda$'s in Eq.~(\ref{eq:Hilbertdecomp}) which can be uniquely specified by eigenvalues under a minimal set of generators of the center $\mZ$.
Further, for each $\lambda$, there are $d_\lambda$ number of identical $D_\lambda$-dimensional Krylov subspaces, which can be uniquely labelled by eigenvalues under a minimal set of generators of any maximal Abelian subalgebra of $\mC$~\cite{moudgalya2021hilbert}.
It is also possible that $D_\lambda = 1$ for some $\lambda$, which correspond to ``singlets" of the algebra $\mA$~\cite{moudgalya2022from}, i.e., simultaneous eigenstates of all the operators in the algebra $\mA$, including the family of Hamiltonians we are interested in. 
The singlets can either be \textit{degenerate} or \textit{non-degenerate} depending on whether the corresponding $d_\lambda$ is greater than $1$ or equal to $1$ respectively, and they play an important role in the study of quantum many-body scars~\cite{moudgalya2022exhaustive}.
All singlets have the property that any ``ket-bra" operator of the form $\ketbra{\psi_{\lambda,\alpha}}{\psi_{\lambda,\beta}}$ for any two degenerate singlets $\ket{\psi_{\lambda,\alpha}}$ and $\ket{\psi_{\lambda,\beta}}$, not necessarily distinct, are a part of the commutant algebra $\mC$.
With this background, in the following sections, we address the following question: Given a family of Hamiltonians of the form of Eq.~(\ref{eq:genhamil}) or a local algebra $\mA = \lgen \{\hH_\alpha\}\rgen$, how does one numerically determine the exhaustive list or number of operators in its commutant $\mC$, construct the decomposition of Eq.~(\ref{eq:Ablock}), and determine the associated dimensions $\{D_\lambda\}$ and $\{d_\lambda\}$? 
We discuss two numerical methods that answer (parts of) this question, and the best method depends on the system being studied.
\section{Simultaneous Block Diagonalization}\label{sec:commutantblock}
We first discuss a method to determine the blocks or Krylov subspaces of a given family of Hamiltonians by simultaneously block-diagonalizing the operators $\{\hH_\alpha\}$. 
This effectively implements the unitary transformation of Eqs.~(\ref{eq:Ablock})-(\ref{eq:Zblock}), which gives a direct access to the dimensions $\{D_\lambda\}$ and $\{d_\lambda\}$ and also to the operators in the commutant $\mC$. 
Similar methods have been used in the literature to ``detect" symmetries in one-parameter families of Hamiltonians~\cite{chertkov2020engineering} or unitary operators~\cite{nie2021operator}, to construct noise commutants in the context of quantum error correction~\cite{holbrook2003noiseless, choi2006method, wang2013numerical},  or more generally, to simultaneously block-diagonalize certain sets of operators~\cite{murota2010numerical, maehara2010numerical,bondarenko2021constructing}.
To determine the blocks, it is typically sufficient to work with two randomly chosen Hermitian operators from the family (i.e., from the algebra $\mA = \lgen \{\hH_\alpha\} \rgen$) which we refer to as
\begin{equation}
    H^{(1)} = \sumal{\alpha}{}{J^{(1)}_\alpha \hH_\alpha},\;\;\;H^{(2)} = \sumal{\alpha}{}{J^{(2)}_\alpha \hH_\alpha}.
\label{eq:genericchoice}
\end{equation}
The rationale for working with just two operators is that these two ``randomly chosen" operators typically generate the full algebra $\mA$, i.e., $\mA = \lgen H^{(1)}, H^{(2)} \rgen$.
However, we emphasize that $\hH_\alpha$ here must be ``generic enough" for this to happen and for the procedure we outline below to work flawlessly, and we do not want them to have any other degeneracies apart from those due to the structure in Eq.~(\ref{eq:Ablock}).
An example of such ``accidental" degeneracies occurs when all the $\{\hH_\alpha\}$ are non-interacting particle-hole symmetric terms (e.g., those shown in \#2 in Tab.~II or III in \cite{moudgalya2022from}), in which case $H^{(1)}$ and $H^{(2)}$ have additional degeneracies due to particle-hole symmetry in the single-particle spectrum, which are not explained by the commutant language. 
We can circumvent such issues by including (Hermitianized) products of the original generators $\{\hH_\alpha\}$ to the list of generators, e.g., operators such as $i[\hH_\alpha, \hH_\beta]$ (which would still be a non-interacting Hamiltonian but typically without the particle-hole-like symmetries) or $\{\hH_\alpha, \hH_\beta\}$ where $\{\cdot\}$ denotes the anticommutator (which would typically be an interacting Hamiltonian). 
Once we have chosen such generic operators, we start by diagonalizing $H^{(1)}$, and expressing $H^{(2)}$ in the eigenbasis of $H^{(1)}$.
\subsection{Extracting \texorpdfstring{$\mZ$}{}}\label{subsec:Zextract}
We start with extracting operators in the center $\mZ$, and to do this, it is sufficient to identify blocks labelled by different $\lambda$.
It is clear from Eq.~(\ref{eq:Zblock}) that the projectors onto these blocks span the center $\mZ$. 
Since all operators in the center $\mZ$ commute with $H^{(1)}$, the eigenstates of $H^{(1)}$ can be labelled by (i.e., have definite) eigenvalues under operators in (or generators of) $\mZ$.\footnote{Since we assume that the generic (random) $H^{(1)}$ does not show any ``accidental" degeneracies, i.e., the corresponding matrices $\{M^{\lambda}(H^{(1)})\}$ are some generic matrices unrelated to each other for different $\lambda$'s, any set of degenerate eigenvectors of $H^{(1)}$ must belong to precisely one $D_\lambda d_\lambda$-dimensional block $\lambda$ and hence these states must be eigenvectors of all operators in $\mZ$.
}
Since all operators in $\mZ$ also commute with $H^{(2)}$, its matrix elements between eigenstates of $H^{(1)}$ that differ in eigenvalues under some operators in $\mZ$ vanish.  
Using the form of the operators in $\mZ$ of Eq.~(\ref{eq:Zblock}), we can conclude that the matrix of $H^{(2)}$ in the eigenbasis of $H^{(1)}$ is certainly block-diagonal with blocks of sizes given by $\{D_\lambda d_\lambda\}$, which are all labelled by different $\lambda$, but this does not rule out the existence of smaller blocks within each of these blocks.
Nevertheless, for generic choices of $H^{(1)}$ and $H^{(2)}$, we can rule out the existence of smaller blocks within blocks in $H^{(2)}$ that are two-dimensional or more.
As evident from Eq.~(\ref{eq:Ablock}), when restricted to a block labelled by $\lambda$, there exists a unitary operator that transforms $H^{(1)}$ and $H^{(2)}$ into $M^{(1)}_{D_\lambda} \otimes \mathds{1}_{d_\lambda}$ and $M^{(2)}_{D_\lambda} \otimes \mathds{1}_{d_\lambda}$ respectively [where $M^{(1)}_{D_\lambda} \defn M^\lambda(H^{(1)})$ and $M^{(2)}_{D_\lambda} \defn M^\lambda(H^{(2)})$].
For $D_\lambda \geq 2$, it is easy to check that for generic choices of $M^{(2)}_{D_\lambda}$ and $M^{(1)}_{D_\lambda}$, the matrix of $M^{(2)}_{D_\lambda} \otimes \mathds{1}_{d_\lambda}$ in a ``random" eigenbasis\footnote{\label{foot:randomize}By this we mean that eigenvectors within degenerate subspaces are ``randomly" chosen.
In practice, this is not guaranteed to be the case for black-box eigensolvers, i.e., the eigenvectors within degenerate subspaces might have some hidden structure (e.g., they are sometimes organized according to the total spin in the computational basis).
Hence it is better to explicitly act with a random unitary to randomize the eigenvectors within each degenerate subspace to ensure that this method works.} of $M^{(1)}_{D_\lambda} \otimes \mathds{1}_{d_\lambda}$ (hence the matrix of $H^{(2)}$ in the eigenbasis of $H^{(1)}$ restricted to the block labelled by $\lambda$) is not expected to have any smaller block-diagonal structure, allowing us to determine these blocks by expressing $H^{(2)}$ in the eigenbasis of $H^{(1)}$. 
On the other hand, this does not identify the blocks where $D_\lambda = 1$ (i.e., blocks composed of singlets of the algebra $\mA$), since both $H^{(1)}$ and $H^{(2)}$ restricted to such blocks are proportional to the identity $(\mathds{1}_{D_\lambda = 1} \otimes \mathds{1}_{d_\lambda}) = \mathds{1}_{d_\lambda}$, hence $H^{(2)}$ is diagonal in the eigenbasis of $H^{(1)}$.
However, the singlet blocks labelled by different $\lambda$ are generically non-degenerate under eigenvalues of $H^{(1)}$ and $H^{(2)}$, hence each such degenerate subspace of $H^{(1)}$ and $H^{(2)}$ corresponds to a block labelled by $\lambda$ consisting of $d_\lambda$ degenerate singlets of $\mA$. 
This completes the identification of the blocks labelled by distinct $\lambda$ in Eq.~(\ref{eq:Hilbertdecomp}), and an orthogonal basis (with respect to the Frobenius inner product) for the center $\mZ$ is given by the projectors onto these blocks.
When $\mC$ is Abelian, these projectors span the full commutant since $\mC = \mZ$.
Hence this procedure is sufficient for constructing Abelian commutants and identifying a subset of Abelian symmetries of a given family of systems, which was demonstrated in \cite{chertkov2020engineering, nie2021operator}.
\subsection{Extracting \texorpdfstring{$\{D_\lambda\}$}{} and \texorpdfstring{$\{d_\lambda\}$}{}}\label{subsec:Ddextract}
Once the blocks labelled by $\lambda$ are identified, the dimensions $\{D_\lambda\}$ and $\{d_\lambda\}$ can directly be extracted as follows. 
For random choices of the coefficients, the eigenvalues of $H^{(1)}$ and $H^{(2)}$ restricted to a block labelled by $\lambda$ appear in multiplets of degeneracies given by $\{d_\lambda\}$, since they can both be expressed as shown in Eq.~(\ref{eq:Ablock}). 
These multiplicities of the eigenvalues of $H^{(2)}$ within each $D_\lambda d_\lambda$-dimensional block can be used to determine $d_\lambda$ corresponding to that block, which can in turn be used to deduce $D_\lambda$.
This allows us to completely determine the dimensions $\{D_\lambda\}$ and $\{d_\lambda\}$, starting from two generic choices of operators $H^{(1)}$ and $H^{(2)}$. 
Quantities such as the number of Krylov subspaces, the dimensions of (i.e., number of linearly independent operators in) the commutant and the local algebra can then be computed directly: they are given by $K = \sum_\lambda{d_\lambda}$, $\dim(\mC) = \sum_\lambda{d_\lambda^2}$, and $\dim(\mA) = \sum_\lambda{D_\lambda^2}$ respectively~\cite{moudgalya2021hilbert}.
Note that while this extraction of the data $\{D_\lambda\}, \{d_\lambda\}$ works in general, we do not yet have the information about the basis that gives the subblock-diagonal structure within each block $\lambda$, Eqs.~(\ref{eq:Ablock})-(\ref{eq:Cblock}).
\subsection{Extracting \texorpdfstring{$\mC$}{} for Non-Abelian Commutants}\label{subsec:nonabelianC}
While the method described in Sec.~\ref{subsec:Zextract} completes the simultaneous block diagonalization and the construction of the full commutant when $\mC$ is Abelian, 
there are additional operators in the commutant when $\mC$ is non-Abelian, and more work is needed.
Let us denote the tensored basis inside the block $\lambda$ implied in the Hilbert space decompositions in  Eqs.~(\ref{eq:Hilbertdecomp})-(\ref{eq:Zblock}) as $\{\ket{\psi_{\alpha\beta}}\}$, $1 \leq \alpha \leq D_\lambda$, $1 \leq \beta \leq d_\lambda$.
For a $D_\lambda$-dimensional Krylov subspace $\mK_{\beta} = \textrm{span}_\alpha\{\ket{\psi_{\alpha\beta}}\}$, we can construct the projectors onto the Krylov subspaces $\Pi_{\beta,\beta} \defn \sum_\alpha{\ketbra{\psi_{\alpha\beta}}{\psi_{\alpha\beta}}}$.
In addition, for degenerate Krylov subspaces $\mK_\beta$ and $\mK_{\beta'}$ for $\beta \neq \beta'$ [i.e., distinct blocks labelled by the same $\lambda$ in Eq.~(\ref{eq:Ablock})], we can construct the operators $\Pi_{\beta,\beta'} \defn \sum_\alpha{\ketbra{\psi_{\alpha\beta}}{\psi_{\alpha\beta'}}}$. 
It is then easy to show that the operators $\{\Pi_{\beta,\beta'}\}$ span the full commutant~\cite{moudgalya2021hilbert}. 
In App.~\ref{app:nonabelianblock} we discuss a method for constructing the operators of the commutant using the matrix elements of $H^{(2)}$ in the eigenbasis of $H^{(1)}$ restricted to a block labelled by $\lambda$ with $D_\lambda \geq 2$.
In particular, we use the fact that within such a block $\lambda$, $H^{(1)}$ and $H^{(2)}$ can be unitarily transformed into $M^{(1)}_{D_\lambda} \otimes \mathds{1}_{d_\lambda}$ and $M^{(2)}_{D_\lambda} \otimes \mathds{1}_{d_\lambda}$, and we assume that there are no degeneracies in the spectrum of $M^{(1)}_{D_\lambda}$ for a random choice of $H^{(1)}$.
This allows us to express the operators in the commutant restricted to the block $\lambda$ in terms of known quantities [see Eq.~(\ref{eq:Pitildfinal})] and hence construct the full commutant.
For singlet blocks with $D_\lambda = 1$, the operators in the commutant are simply the ``ket-bra" operators of the degenerate singlets (e.g., $\ketbra{\psi_\beta}{\psi_{\beta'}}$ for degenerate singlets $\ket{\psi_\beta}$ and $\ket{\psi_{\beta'}}$) which can be constructed directly from the corresponding eigenvectors of $H^{(2)}$ (which are also eigenvectors of $H^{(1)}$).
Note that this method requires a full exact diagonalization of $H^{(1)}$, hence its time complexity grows exponentially with system size [as $\mO(d^{3L}_{\loc})$ for a system of size $L$ and local Hilbert space dimension $d_{\loc}$], and its application is practical only for small system sizes.  
\subsection{Examples}\label{subsec:blockdiagexamples}
\begin{figure*}[t!]
\centering
\includegraphics[scale = 1]{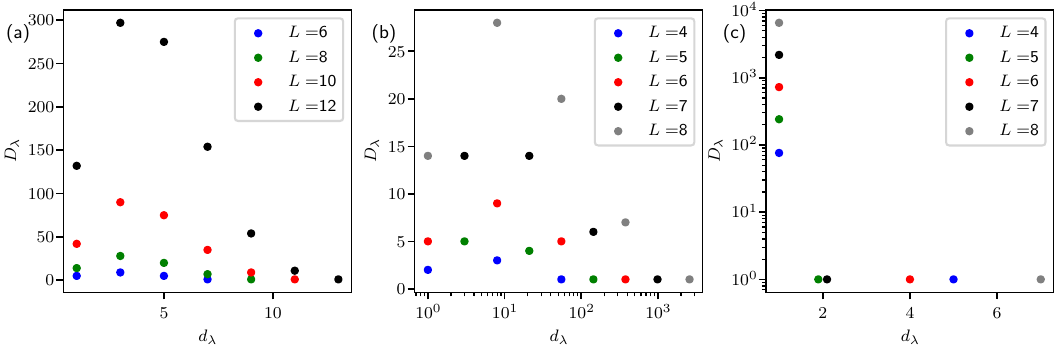} 
\caption{
(Color online) The sizes and degeneracies of the blocks (Krylov subspaces) for several types of Hamiltonians with non-trivial commutant algebras, extracted using the Simultaneous Block Diagonalization method for various system sizes $L$. 
$\{D_\lambda\}$ denote the sizes of the blocks and $\{d_\lambda\}$ denote their degeneracies.
For details in each case, see Sec.~\ref{subsec:blockdiagexamples}.
(a) Conventional Symmetries: Spin-1/2 Heisenberg model.
The blocks are the $SU(2)$ total spin quantum number sectors.
(b) Hilbert Space Fragmentation: Spin-1 Temperley-Lieb Models~\cite{moudgalya2021hilbert}.
The small sizes of blocks and large degeneracies indicate the presence of Quantum Hilbert Space Fragmentation.
(c) Quantum Many-Body Scars (QMBS): Spin-1 models realizing the periodic boundary condition spin-1 AKLT tower of scars~\cite{moudgalya2018exact,moudgalya2018entanglement} as degenerate QMBS~\cite{moudgalya2022exhaustive}.
A few one-dimensional blocks along with one outstanding large block indicate the presence of QMBS.
Note that for the sake of clarity any ``degeneracies" in $(D_\lambda, d_\lambda)$ have been resolved by introducing a small horizontal separation.}
\label{fig:simulblockdiag}
\end{figure*}
We now discuss some examples of this method being applied to extract values of $\{D_\lambda\}$ and $\{d_\lambda\}$ in various types of systems.
We separately consider models with conventional symmetries, those with Hilbert space fragmentation, and those with QMBS, and we depict all the results in Figs.~\ref{fig:simulblockdiag}(a)-(c).
\subsubsection{Conventional Symmetries}
We start with an example of conventional symmetries in the case of the family of one-dimensional spin-1/2 Heisenberg models, given by the family of Hamiltonians $H = \sum_{j = 1}^{L}{J_j (\vec{S}_j \cdot \vec{S}_{j+1})}$, where $\vec{S}_j = (S^x_j, S^y_j, S^z_j)$ are the spin-1/2 operators on site $j$.
The boundary conditions for these can be either periodic or open, and the symmetry algebra does not depend on this choice.  
This family of systems has been discussed using the commutant algebra framework in \cite{moudgalya2021hilbert, moudgalya2022from}, and the local algebra and commutant algebra pair is given by
\begin{equation}
    \mA_{SU(2)} = \lgen \{\vec{S}_j \cdot \vec{S}_{j+1}\} \rgen,\;\;\mC_{SU(2)} = \lgen S^x_{\tot}, S^y_{\tot}, S^z_{\tot} \rgen,
\label{eq:Heisalgebrapair}
\end{equation}
where $S^\alpha_{\tot} \defn \sum_j{S^\alpha_j}$, $\alpha \in \{x, y, z\}$, is the total spin operator in the direction $\hat{\alpha}$. 
The dimensions of irreducible representations for this pair of algebras, i.e., the numbers $\{D_\lambda\}$ and $\{d_\lambda\}$, obtained using the simultaneous block-diagonalization method are shown in Fig.~\ref{fig:simulblockdiag}(a).   
For this pair of algebras, the distinct $\lambda$ in the decompositions of Eqs.~(\ref{eq:Ablock})-(\ref{eq:Zblock}) correspond to the total spin quantum number which for even system size $L$ takes values $\lambda = 0, 1, \dots, L/2$ and for odd system size takes $\lambda = 1/2, 3/2, \dots, L/2$ [namely, the eigenvalues of $\vec{S}^2_{\tot}$ are given by $\lambda (\lambda + 1)$].
The horizontal axis in the figure indicates $d_\lambda = 2\lambda + 1$ which is the familiar degeneracy of spin-$\lambda$ sector.
In particular, $d_{\lambda=0} = 1$ marks the space of spin-singlets (i.e., states that have eigenvalue $0$ under $S^2_{\tot}$), while $d_{\lambda=L/2} = L+1$ marks the ferromagnetic manifold.
Note that all ferromagnetic states are degenerate singlets of the algebra $\mA_{SU(2)}$ generated by the individual Heisenberg terms, hence $D_{\lambda=L/2} = 1$.
In general, the sizes of each of the quantum number sectors of spin-$\lambda$, given by $D_\lambda$, are known exactly~\cite{readsaleur2007, moudgalya2021hilbert}
\begin{equation}
D_\lambda = \binom{L}{L/2+\lambda} - \binom{L}{L/2+\lambda+1}.
\label{eq:Dlam}
\end{equation}
As we depict in Fig.~\ref{fig:simulblockdiag}(a), the simultaneous block diagonalization procedure reproduced all these numbers correctly.
While the blocks labelled by different values of $\lambda$, hence the operators in the center $\mZ_{SU(2)} \defn \mA_{SU(2)} \cap \mC_{SU(2)}$, were directly obtained using the method described in Sec.~\ref{subsec:Zextract}, extracting the full non-Abelian commutant $\mC_{SU(2)}$ required the implementation of the procedure described in Sec.~\ref{subsec:nonabelianC}.
To ensure that these procedures work as intended, we found that the explicit randomization of eigenvectors, described in footnote~\ref{foot:randomize} was particularly important. 
\subsubsection{Hilbert Space Fragmentation}\label{subsubsec:blockdiagfrag}
We move on to an example of this method applied to models exhibiting Hilbert space fragmentation. 
We focus on the spin-1 Temperley-Lieb models, given by the family of Hamiltonians $H_{TL} = \sum_{j = 1}^{L-1}{J_j (\vec{S}_j \cdot \vec{S}_{j+1})^2}$, where $\vec{S}_j = (S^x_j, S^y_j, S^z_j)$ here are the spin-1 matrices on site $j$, and we have used open boundary conditions. 
The local algebra in this case is given by 
\begin{equation}
    \mA_{\text{TL}} = \lgen \{(\vec{S}_j\cdot\vec{S}_{j+1})^2\} \rgen,
\label{eq:TLlocalalg}
\end{equation}
and the corresponding commutant $\mC_{\text{TL}}$ was computed explicitly in \cite{readsaleur2007, moudgalya2021hilbert}; we are not aware of a compact expression for $\mC_{TL}$.
The values of $\{d_\lambda\}$ and $\{D_\lambda\}$ for this pair of algebras is shown in Fig.~\ref{fig:simulblockdiag}(b).
These numbers were also analytically computed in \cite{readsaleur2007}, and while the number of distinct blocks labelled by distinct $\lambda$ and the values $D_\lambda$ are identical to the spin-1/2 Heisenberg model case in the previous section, the degeneracies $\{d_\lambda\}$ are given by
\begin{equation}
    d_\lambda = [2\lambda + 1]_q \defn \frac{q^{2\lambda + 1} - q^{-(2\lambda+1)}}{q - q^{-1}},\;\;q = \frac{3 + \sqrt{5}}{2}.
\label{eq:dlamq}
\end{equation}
For example, the block with the largest $d_\lambda$, given by the block labelled by $\lambda = L/2$, also has $D_{\lambda = L/2} = 1$ and contains ``ferromagnetic" product states $\ket{\alpha_1,\alpha_2,\dots,\alpha_L}$ with $\alpha_j \neq \alpha_{j+1}, j=1,\dots,L-1$ for OBC along with their full $SU(3)$ lowered multiplets, see \cite{readsaleur2007} and references therein for the details; hence $d_{\lambda = L/2}$ grows exponentially with system size $L$. 
All our numerical results shown in Fig.~\ref{fig:simulblockdiag}(b) are consistent with the above analytical results~\cite{readsaleur2007} and provide an independent check of these predictions.
\subsubsection{Quantum Many-Body Scars}
Finally, we provide an example of this method applied to a model of QMBS. 
As a non-trivial example, we consider spin-1 models on a periodic chain that realize the exact tower of scars found in the AKLT model~\cite{moudgalya2018exact, moudgalya2018entanglement, mark2020unified, moudgalya2020large, odea2020from} as exact degenerate scars~\cite{moudgalya2022exhaustive}.
The local algebra for this case is given by~\cite{moudgalya2022exhaustive}
\begin{equation}
    \tmA^{(p)}_{\scar} = \lgen \{\Pi_{[j,j+2]} h_{[j,j+2]} \Pi_{[j,j+2]}\} \rgen,
\label{eq:PBCAKLTlocalalg}
\end{equation}
where $h_{[j,j+2]}$ is a sufficiently generic three-site operator, and $\Pi_{[j,j+2]}$ are three-site projectors chosen such that they vanish on the AKLT scar tower, see App.~D of \cite{moudgalya2022exhaustive} for details on their precise construction.
As conjectured in \cite{moudgalya2022exhaustive}, the commutant algebra $\tmC^{(p)}_{\scar}$ is expected to be spanned by ket-bra operators of the $N_{\scar}$ states in the common kernel of the projectors $\{\Pi_{[j,j+2]}\}$, where $N_{\scar}$ depends on the system size $L$.
Equivalently, we expect the Hilbert space decomposition in Eqs.~(\ref{eq:Ablock})-(\ref{eq:Zblock}) to simply consists of two distinct $\lambda$'s, one corresponding to the thermal block and one to the scar block, which we denote by $\lambda = \tt{t}$ and $\lambda = \tt{s}$ respectively.
We then expect the scar block to have $(D_{\lambda = \tt{s}}, d_{\lambda = \tt{s}}) = (1, N_{\scar})$, and the thermal block to have $(D_{\lambda = \tt{t}}, d_{\lambda = \tt{t}}) = (\dim(\mH) - N_{\scar}, 1)$, where $\dim(\mH) = 3^L$ is the total dimension of the Hilbert space.
The common kernel of these projectors for various system sizes has been conjectured in \cite{moudgalya2022exhaustive} (see Sec.~V C and App.~D there), and we briefly summarize the results here. 
For $L$ odd, the common kernel is spanned by two states -- the AKLT ground state and the spin-polarized ferromagnetic state, hence $d_{\lambda = s} = 2$. 
For $L = 2 \times \text{odd}$, the kernel is spanned by the $L/2$ QMBS tower of states, denoted by $\{\ket{\psi_n}\}$ and defined in Eq.~(22) of \cite{moudgalya2022exhaustive}, as well as the ferromagnetic state, which is not a part of the tower of QMBS for these system sizes~\cite{mark2020unified}; hence the total number of states in the kernel is $L/2 + 1$.  
On the other hand, for $L = 2 \times \text{even}$, in addition to the $L/2 + 1$ QMBS tower of states $\{\ket{\psi_n}\}$ (which now includes the ferromagnetic state), there are two additional states in the kernel, denoted by $\ket{1_{k = \pm \pi/2}}$ in \cite{moudgalya2018exact, moudgalya2022exhaustive}, which add up to $L/2 + 3$ states in the kernel.
These explain all the values of $\{D_\lambda\}$ and $\{d_\lambda\}$ shown in Fig.~\ref{fig:simulblockdiag}(c) for various system sizes. 
Note that these results were only conjectured and not proven in our previous work \cite{moudgalya2022exhaustive}, hence the results here illustrate non-trivial discovery/validation of the full commutant algebras.
\section{Liouvillian Approach}\label{sec:commutantLiouv}
We now discuss an alternate method to construct operators in and determine the dimension of the commutant, which, in certain cases, allows us to determine the commutant for much larger system sizes. 
We start by interpreting operators $\hO$ as vectors $\oket{\hO}$, hence we obtain
\begin{equation}
    [\hH_\alpha, \hO] = 0\;\;\iff\;\;\overbrace{\left(\hH_\alpha \otimes \mathds{1} - \mathds{1} \otimes \hH_\alpha^T\right)}^{\hmL_{\hH_\alpha} \defn}\oket{\hO} = 0,
\label{eq:commliouv}
\end{equation}
where $\hmL_{\hH_\alpha}$ is the Liouvillian corresponding to the term $\hH_\alpha$, i.e., it represents the adjoint action of the Hamiltonian, hence $\hmL_{\hH_\alpha}\opket{\bullet} \defn [\hH_\alpha, \bullet]$. 
Using Eq.~(\ref{eq:commliouv}) and the definition of the commutant in Eq.~(\ref{eq:eachtermcommute}), the commutant is the common kernel of the Liouvillian superoperators $\{\hmL_{\hH_\alpha}\}$. 
\subsection{Mapping onto a Frustration-Free Ground State Problem}\label{subsec:LiouvFF}
This common kernel can be expressed as the null subspace of the positive semi-definite (p.s.d.) superoperator
\begin{equation}
    \hmP \defn \sumal{\alpha}{}{\overbrace{\hmL_{\hH_\alpha}^\dagger \hmL_{\hH_\alpha}}^{\hmP_{\hH_\alpha} \defn}},\;\;\;\hmP\oket{\hO} = 0 \iff \hmL_{\hH_\alpha}\oket{\hO} = 0\;\;\forall \alpha, 
\label{eq:psdLiouv}
\end{equation}
where the second condition follows due the p.s.d.\ property of all $\hmP_{\hH_\alpha}$. 
The dimension of the commutant, $\textrm{dim}(\mC)$, is simply the dimension of this nullspace.
The commutant can be numerically computed straightforwardly using this approach, although the time-complexity is worse than the previous simultaneous block-diagonalization method since it involves the diagonalization of the superoperator $\hmP$ and scales as $\mathcal{O}(d^{6L}_{\loc})$.
Hence a direct application of this method is usually limited to extremely small system sizes, although efficient methods for determining the kernel (e.g., Lanczos algorithm) can be used to improve its performance. 
Nevertheless, further progress can be made in some cases by noting that the operators $\{\hH_\alpha\}$ are local terms, either strictly local or extensive local terms. 
When they are strictly local, so are the Hermitian p.s.d.\ superoperators $\hmP_{\hH_\alpha} \defn \hmL_{\hH_\alpha}^\dagger \hmL_{\hH_\alpha}$.\footnote{Concretely, we can interpret the two tensored copies of the Hilbert space in Eq.~(\ref{eq:commliouv}) as two legs of a ladder (or two layers of a bilayer in higher dimensions), as common in the study of superoperators, e.g., in the Lindblad master equation~\cite{ziolkowska2020yangbaxter}. 
The superoperators and operators in the original Hilbert space are operators and states in the doubled (ladder) Hilbert space respectively.
It is then clear that if $\hH_\alpha$ is strictly local in the original Hilbert space, so are $\hmL_{\hH_\alpha}$ and also $\hmP_{\hH_\alpha}$ in the doubled (ladder) Hilbert space.}
Hence operators in the commutant are the frustration-free ``ground states" of a local superoperator ``Hamiltonian" $\hmP$.
\subsection{Efficient Method in One Dimension}\label{subsec:1deff}
\begin{figure*}
\centering
\includegraphics[scale=0.5]{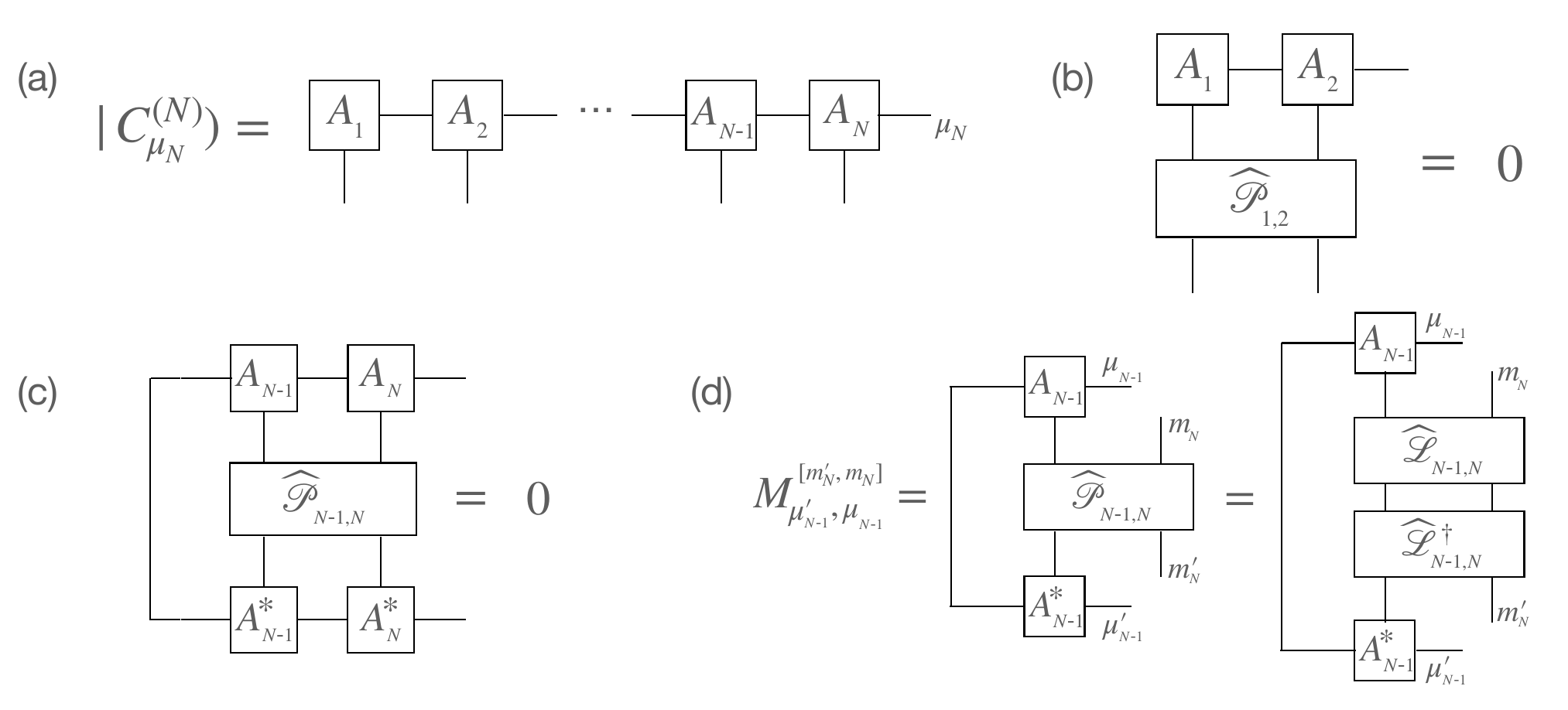}
\caption{
Liouvillian approach for the efficient construction of commutants of nearest-neighbor bond algebras in one dimension with open boundary conditions. (a) MPS representation for the basis vectors of the commutant on $N$ sites. (b) Equation for solving for the leftmost two unknown tensors $A_1$ and $A_2$ by diagonalizing the known matrix $\hmP_{1,2}$. (c) Equation for recursively solving for the unknown tensor $A_N$ in terms of the known tensor $A_{N-1}$ and the known matrix $\hmP_{N-1, N}$. (d) Matrix $M$ for which the tensor $A_N$ is an eigenvector with eigenvalue $0$. }
\label{fig:MPScomm}
\end{figure*}
Efficient numerical methods to obtain the frustration-free ground states of a one-dimensional Hamiltonian are known in the literature~\cite{debeaudrap2010solving, movassagh2010unfrustrated, yao2022bounding}, and they can be directly adapted to be applied to this problem.
We discuss one such method in detail in App.~\ref{app:efficientcommutant1d} and provide a summary here.
Note that this method borrows ideas from existing numerical algorithms based on Matrix Product States (MPS), particularly the Density Matrix Renormalization Group (DMRG)~\cite{schollwock2011density}.
However, unlike DMRG, this algorithm is ``exact", i.e., it does not use any variational optimizations or approximations (hence its accuracy is limited only by accumulations of machine round-off errors).
\subsubsection{Strictly local generators with OBC}
We start with the simplest case, namely, obtaining commutants of bond algebras generated by nearest-neighbor terms, say $\{\hH_{j,j+1}\}$, with open boundary conditions (OBC). 
We can then define the nearest-neighbor Liouvillians $\{\hmL_{j,j+1}\}$ and correspondingly $\{\hmP_{j,j+1}\}$ using Eqs.~(\ref{eq:commliouv}) and (\ref{eq:psdLiouv}) respectively.
For a system of size $L$, we start by envisioning the $\chi_L \defn \dim(\mC_L)$ operators in the commutant $\mC_L$, spanned by an orthonormal basis $\{\oket{C^{(L)}_{\mu_L}}\}_{\mu_L = 1}^{\chi_L}$, as a Matrix Product State (MPS) consisting of tensors $\{A_N\}_{N = 1}^L$ with an open auxiliary index on the right.
We wish to construct these tensors such that this property is true for any system of size $2 \leq N \leq L$, i.e., the $N$-site MPS with an open auxiliary index on the right [see Eq.~(\ref{eq:CNMPSform}) and Fig.~\ref{fig:MPScomm}(a)] spans the $N$-site commutant $\mC_N$ (i.e., the centralizer of the operators $\{\hH_{j,j+1}\}_{j = 1}^{N-1}$).
Hence the MPS is not a standard one, and its bond dimension can grow from left to right, and the $N$-th bond from the left, $N = 2, \cdots, L$, has a dimension $\chi_N \defn \dim(\mC_N)$.
We then solve for the tensors of the MPS recursively from the left using ideas similar to DMRG~\cite{schollwock2011density}. 
In particular, we solve for the two tensors $A_1$ and $A_2$ on the left by requiring that the MPS restricted to two sites is annihilated by $\hmP_{1,2}$ [see Eq.~(\ref{eq:twositeschmidt}) and Fig.~\ref{fig:MPScomm}(b)]. 
We then determine the rest of the tensors $\{A_N\}_{N = 3}^L$ recursively, i.e., we can use the form of the $(N-1)$-th tensor to determine the $N$-th by requiring that $\hmP_{N-1, N}$ annihilates the $N$-site MPS [see Fig.~\ref{fig:MPScomm}(c)]; in particular, this condition can be rephrased in terms of determining the nullspace of an $d^2_{\loc}\chi_{N-1}$-dimensional matrix $M$ [see Eq.~(\ref{eq:effectivematrix}) and Fig.~\ref{fig:MPScomm}(d)].
The dimension of the auxiliary index $\chi_N$ for the $N$-th tensor $A_N$ (which is also the dimension of the $N$-site commutant) is simply the dimension of the nullspace of that matrix.
Hence the dimension of the commutant for a system of $L$ sites can be determined using this recursive process.
As we discuss in App.~\ref{app:efficientcommutant1d}, this method can be generalized in many directions. 
For bond algebras in OBC that are generated by $r$-site strictly local terms, the $N$-th tensor $A_N$ can be recursively determined using the previous $(r-1)$ tensors $\{A_{N - r +1}, \cdots, A_{N-1}\}$, and it requires a diagonalization of an $d^2_{\loc}\chi_{N-1}$-dimensional matrix. 
Hence the time-complexity of this method for such a bond algebra and large systems of size $L$ naively scales as $\mathcal{O}(L d^{6}_{\loc} \chi^3_{L})$, although in practice there can be steps in the computation that are comparable in time-complexity, e.g., the construction of the aforementioned matrix itself takes $\mathcal{O}(d^{4(r-1)}_{\loc}\chi^3_{L})$.
Nevertheless, these estimates make it clear that the efficiency of this method depends on the scaling of the commutant dimension with system size. 
If the commutant dimension stays constant or grows at most polynomially with the system size, e.g., in systems with conventional symmetries such as $U(1)$ or $SU(2)$~\cite{moudgalya2021hilbert, moudgalya2022from} or systems exhibiting quantum many-body scars~\cite{moudgalya2022exhaustive}, this method can be highly efficient, as we show in Figs.~\ref{fig:liouvcomm}(a) and \ref{fig:liouvcomm}(c). 
However, if the commutant dimension grows exponentially with the system size, e.g., for systems exhibiting Hilbert space fragmentation~\cite{moudgalya2021hilbert}, the accessible system sizes are limited, as we show in Fig.~\ref{fig:liouvcomm}(b).
In such cases, it can in practice be more efficient to work with the simultaneous block-diagonalization method discussed in Sec.~\ref{sec:commutantblock}. 
\subsubsection{Strictly local generators with PBC or extensive local generators}
While this MPS construction works rather neatly for commutants of bond algebras generated by strictly local terms with OBC, additional work needs to be done to obtain commutants of algebras generated with strictly local terms with periodic boundary conditions (PBC) or with extensive local operators. 
In these cases, we first express Eq.~(\ref{eq:psdLiouv}) as $\hmP = \hmP_{\obc} + \hmP_{\strad} + \hmP_{\ext}$, where (i) $\hmP_{\obc}$, (ii) $\hmP_{\strad}$, and (iii) $\hmP_{\ext}$ respectively contain the $\hmP_{\hH_\alpha}$'s corresponding to the generators $\hH_\alpha$ that are (i) strictly local and can be viewed as generating the OBC bond algebra, (ii) strictly local and straddle the PBC ``boundary", and (iii) extensive local.
We then proceed by first determining the $\chi_L$-dimensional OBC commutant, i.e., the kernel of $\hmP_{\obc}$, and we denote this subspace as $\mC_{L} \defn \{\oket{C^{(L)}_{\mu_L}}\}_{\mu_L = 1}^{\chi_L}$. 
We then compute $\hmT_{\strad/\ext}$, the $\chi_L$-dimensional matrices that are the restrictions of $\hmP_{\strad}$ and $\hmP_{\ext}$ in the subspace $\mC_L$; their matrix elements are given by
\begin{equation}
    (\hmT_{\strad/\ext})_{\mu_L, \mu'_L} \defn \obra{C^{(L)}_{\mu_L}} \hmP_{\strad/\ext} \oket{C^{(L)}_{\mu'_L}}.
\label{eq:Trestmatrix}
\end{equation}
The kernel of the $\chi_L$-dimensional matrix $\hmT_{\strad} + \hmT_{\ext}$ is then the commutant of the full local algebra, including all the generators.\footnote{Note that a zero eigenvector of $\hmT_{\strad/\ext}$, i.e., $\hmP_{\strad/\ext}$ restricted to the space $\mC_L$, is always a zero eigenvector of $\hmP_{\strad/\ext}$. This is because for any such vector $\oket{\hO_0} \in \mC_L$, we have by definition $\obra{\hO_0} \hmP_{\strad/\ext} \oket{\hO_0} = 0$. Since $\hmP_{\strad/\ext}$ is positive semi-definite, this means that $\hmP_{\strad/\ext}\oket{\hO_0} = 0$.}
As we discuss in Apps.~\ref{subsec:PBCbondalgebra} and \ref{subsec:localalgebra}, the restricted matrices $\hmT_{\strad}$ and $\hmT_{\ext}$ can be efficiently computed using the MPS form for $\mC_{L}$ and using transfer matrices of its MPS tensors and MPO forms of $\hmP_{\hH_\alpha}$ when $\hH_\alpha$ is extensive local [see Eqs.~(\ref{eq:Tmatrix_via_Es}) and (\ref{eq:Texttransfer})].
Note that this method involves the construction of $\Upsilon\chi^2_{N-1} \times \Upsilon\chi^2_N$ transfer matrices for every $N \leq L$ (where $\Upsilon = 1$ for straddling operators and $\Upsilon \sim \mathcal{O}(1)$ for extensive local operators, related to the bond dimension of the MPO of the corresponding $\hmP_{\hH_\alpha}$), their multiplication, and then the diagonalization a $\chi_L$-dimensional matrix.
Hence the full time-complexity of this method for large system sizes naively scales as $\mathcal{O}(L \Upsilon^2 \chi_L^4)$, although a better scaling might be possible with an efficient tensor contraction ordering~\cite{schindler2020algorithms}.
Although this scaling is polynomial in system size when the OBC commutant dimension scales polynomially, it is worse than the OBC problem, and the large exponents that show up in typical cases of interest (e.g., if $\chi_L \sim L^2$, the time-complexity scales as $\sim L^9$) can be a significant hinderance in practice.
Also note that the time-complexity depends on the scaling of the dimension of the OBC commutant, which might be larger than the scaling of the full commutant when the PBC or extensive local operators are included.
Hence this method might not be efficient even if we expect the full commutant to have a small dimension.
\subsection{Restricting the form of the conserved quantity}\label{subsec:conservedform}
Finally, we show that this method can also be directly extended to search for conserved quantities in the commutant that are of a particular form, e.g., strictly local or extensive local operators of a fixed range. 
We denote the vector space spanned by operators of this form as $\mV$, and a linearly independent basis for $\mV$ by $\{\oket{V_{\mu}}\}$, and construct the operator $\hmP_{\mV}$, defined as the restriction of $\hmP$ to $\mV$.
That is, the matrix elements of $\hmP_{\mV}$ read
\begin{equation}
    (\hmP_{\mV})_{\mu,\mu'} = \obra{V_\mu}\hmP\oket{V_{\mu'}}.
\label{eq:Prestmat}
\end{equation}
Then using the p.s.d.\ property of $\hmP$ we can show that the nullspace of $\hmP_{\mV}$ is spanned by operators in $\mC$ that are in $\mV$, i.e.,
\begin{equation}
    \hmP_{\mV}\oket{\hO} = 0\;\;\iff\;\;\hmP\oket{\hO} = 0\;\;\&\;\;\oket{\hO} \in \mV. 
\label{eq:restrictedop}
\end{equation}
\subsection{Examples}\label{subsec:liouvexamples}
\begin{figure*}[t!]
\centering
\includegraphics[scale = 1]{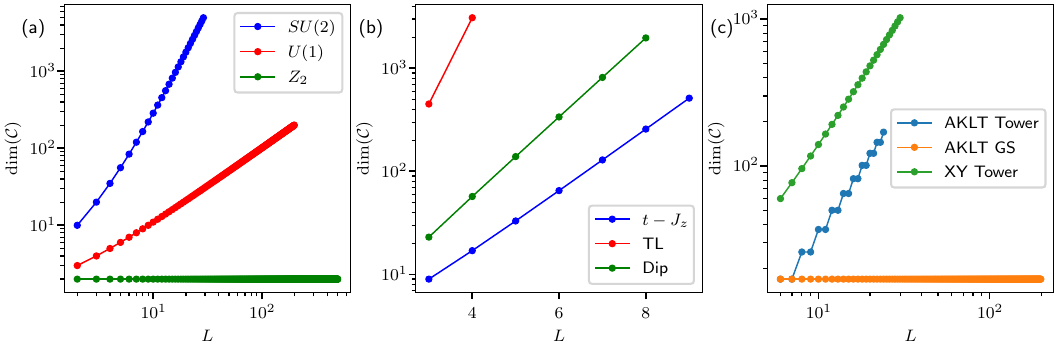} 
\caption{(Color online) The dimension of the commutant algebras for several families of Hamiltonians, extracted using the efficient Liouvillian method.
As discussed in Sec.~\ref{subsec:1deff}, this works best in cases where the local algebra is generated by strictly local terms with open boundary conditions, and we have only demonstrated such cases.
For details in each case, see Sec.~\ref{subsec:liouvexamples}.
(a) Conventional Symmetries: Families of models with $Z_2$, $U(1)$, and $SU(2)$ symmetries; (b) Hilbert Space Fragmentation: $t-J_z$ models, spin-1 Temperley-Lieb (TL) models, spin-1 dipole conserving models; (c) Quantum Many-Body Scars: OBC AKLT ground states as scars, AKLT scar tower as degenerate scars, spin-1 XY $\pi$-bimagnon scar tower as degenerate scars.
Note that $\dim(\mC)$ is a limiting factor in this method, hence smaller system sizes are accessible for systems with larger $\dim(\mC)$.
}
\label{fig:liouvcomm}
\end{figure*}
We now provide examples of computations of the dimension of the commutant algebra, i.e., the number of linearly independent operators in the algebra, using the efficient Liouvillian MPS method.
We again illustrate this separately for cases with regular symmetries, fragmented systems, and QMBS systems, and we depict the results in Fig.~\ref{fig:liouvcomm}(a)-(c).
As discussed in Sec.~\ref{subsec:1deff}, this method is powerful in practice compared to the simultaneous block-diagonalization method only for local algebras generated by strictly local terms with open boundary conditions. 
Note that the dimension of the commutant is related to the dimensions of the irreducible representations as $\dim(\mC) = \sum_{\lambda}{d_\lambda^2}$~\cite{moudgalya2021hilbert}, which provides an alternate way to compute it if the $d_\lambda$'s are known.
\subsubsection{Conventional symmetries}
As examples of systems with conventional symmetries, we consider spin-1/2 systems with $Z_2$, $U(1)$, and $SU(2)$ symmetries. 
The local and commutant algebras for these are discussed in \cite{moudgalya2022from}, see Tab.~I there. 
To summarize, the corresponding algebra pairs are given by
\begin{align}
    &\mA_{Z_2} = \lgen \{S^x_j S^x_{j+1}\}, \{S^z_j\}\rgen, \;\;\; \mC_{Z_2} = \lgen \prod_j{S^z_j} \rgen, \nn \\
    &\mA_{U(1)} = \lgen \{S^x_j S^x_{j+1} + S^y_j S^y_{j+1}\}, \{S^z_j\} \rgen, \;\;\;\mC_{U(1)} = \lgen S^z_{\tot} \rgen, \nn\\
    &\mA_{SU(2)} = \lgen \{\vec{S}_j\cdot\vec{S}_{j+1}\} \rgen, \;\;\;\mC_{SU(2)} = \lgen S^x_{\tot},  S^y_{\tot}, S^z_{\tot}\rgen,
\end{align}
where $S^\alpha_j$, $\alpha \in \{x, y, z\}$ are the spin-1/2 operators on site $j$, and $S^\alpha_{\tot} = \sum_j{S^{\alpha}_j}$.
For the $Z_2$ and $U(1)$ symmetries, the analytically known dimensions of the commutants can be directly obtained by counting the number of linearly independent operators, whereas in the case of $SU(2)$ it is easier to compute $\dim(\mC)$ using the known values of $\{d_\lambda\}$, the dimensions of the irreducible representations. 
In particular, we obtain $\dim(\mC_{Z_2}) = 2$, $\dim(\mC_{U(1)}) = L + 1$, and $\dim(\mC_{SU(2)}) = \binom{L+3}{3}$, and we refer readers to \cite{moudgalya2021hilbert} for the details of the calculations. 
Our numerical results in Fig.~\ref{fig:liouvcomm}(a) are consistent with these analytical predictions. 
Note that the system sizes accessible in the case of $SU(2)$ are much smaller than the others due to the faster scaling of $\dim(\mC_{SU(2)})$ with the system size $L$.
\subsubsection{Hilbert space fragmentation}
For fragmented systems, we apply this method to the $t-J_z$ models~\cite{batista2000tJz, sliom2020}, spin-1 dipole conserving models~\cite{sala2020fragmentation}, and spin-1 TL models~\cite{readsaleur2007}, which have been studied  in the commutant algebra framework in \cite{moudgalya2021hilbert}.
As discussed there, the $t-J_z$ models and the spin-1 dipole conserving models are examples of ``classical fragmentation", i.e., the commutants in these cases, which we denote by $\mC_{t-J_z}$ and $\mC_{\dip}$, are completely spanned by operators that are diagonal in the product states basis.
Hence they are Abelian, and their dimension is simply the number of distinct blocks in the product state basis, i.e., the number of ``Krylov subspaces"~\cite{moudgalya2019thermalization}. 
The structure of the Krylov subspaces in these cases has been discussed in detail in \cite{moudgalya2021hilbert}, and we obtain $\dim(\mC_{t-J_z,\text{obc}}) = 2^{L+1} - 1$ and $\dim(\mC_{\dip,\text{obc}}) = 2 P_{L +1} -1 $, where $P_{L+1}$ is the $(L+1)$th Pell number, which grows exponentially with $L$. 
On the other hand, the spin-1 Temperley-Lieb model exhibits ``quantum fragmentation", and the dimension of the commutant can be computed directly using the dimensions of the irreducible representations, known from earlier literature~\cite{readsaleur2007}, also shown in Eq.~(\ref{eq:dlamq}).
All of these numbers are consistent with the dimensions obtained from the efficient MPS Liouvillian method, shown in Fig.~\ref{fig:liouvcomm}(b).
Note that the system sizes accessible here are much smaller than in the case of conventional symmetries due to the exponential scaling of the commutant dimensions.
\subsubsection{Quantum Many-Body Scars}
Finally, we apply this method to several models exhibiting QMBS studied in \cite{moudgalya2022exhaustive}. 
Note that there is a distinction between systems with degenerate QMBS and non-degenerate QMBS, in particular the local and commutant algebras in these cases are different. 
From the examples studied in \cite{moudgalya2022exhaustive}, it appears that to obtain commutant algebras corresponding to non-degenerate QMBS, it is necessary to include an extensive local operator in the generators of the local algebra, whereas that is not the case for commutant algebras corresponding to degenerate QMBS.  
Since the efficiency of the MPS method is truly evident only for commutants of local algebras generated only by strictly local operators with OBC, we will only consider examples of degenerate QMBS throughout this section. 
First, we consider the case where the four OBC AKLT ground state are degenerate QMBS -- Hamiltonians with this feature can be systematically constructed by an explicit ``embedding" of these states into the spectrum~\cite{shiraishi2017systematic}.
The bond algebra in this case is given by 
\begin{equation}
    \tmA^{\AKLT}_{\scar} = \lgen \{P^{\AKLT}_{j,j+1} h_{j,j+1} P^{\AKLT}_{j,j+1}\} \rgen,
\end{equation}
where $P^{\AKLT}_{j,j+1}$ are the spin-1 AKLT projectors, and $h_{j,j+1}$ is a sufficiently generic two-site operator, see \cite{moudgalya2022exhaustive} for the details. 
The commutant of $\tmA^{\AKLT}_{\scar}$, with OBC, is spanned by the identity operator, along with all ket-bra operators formed by states in the kernel of $\{P^{\AKLT}_{j,j+1}\}$, which are the four OBC AKLT ground states; hence the dimension of the commutant is given by $\dim(\tmC^{\AKLT}_{\scar}) = 17$.
Second, we consider models with the OBC AKLT tower of QMBS~\cite{moudgalya2018exact} as degenerate QMBS, where the bond algebra is given by~\cite{moudgalya2022exhaustive}
\begin{align}
& \tmA^{(o)}_{\scar} = \lgen \Pi^{(l)}_{1, 2} h^{(l)}_{1,2}\Pi^{(l)}_{1, 2},\{\Pi_{[j,j+2]} h_{[j, j+2]} \Pi_{[j,j+2]}\},\nn \\
& \qquad\qquad \Pi^{(r)}_{L-1, L} h^{(r)}_{L-1,L}\Pi^{(r)}_{L-1, L}\rgen,
\label{eq:ACpairobc}
\end{align}
where $\{h_{[j,j+2]}\}$ and $\{h^{(l)}_{1, 2}, h^{(r)}_{L-1, L}\}$ are sufficiently generic operators in the bulk and on the boundary respectively, and $\{\Pi_{[j,j+2]}\}$ and $\{\Pi^{(l)}_{1, 2}$, $\Pi^{(r)}_{L-1, L}\}$ are three-site and two-site bulk and boundary projectors respectively; see App.~D of \cite{moudgalya2022exhaustive} for the details of their construction.
The commutant of this algebra, $\tmC^{(o)}_{\scar}$, was conjectured to be fully spanned by ket-bra operators of the common kernel of the projectors $\{\Pi_{[j,j+2]}\}$ and $\{\Pi^{(l)}_{1, 2}, \Pi^{(r)}_{L-1, L}\}$, along with the identity operator.
Hence we expect its dimension to grow as $N_{\scar}^2 + 1$, where $N_{\scar}$ is the number of states in this kernel. 
As discussed in \cite{moudgalya2022exhaustive} [see Eq.~(D20) there], we expect $N_{\scar} = L/2 + 1$ when $L$ is even, and $N_{\scar} = (L+1)/2$ when $L$ is odd, which then gives the expected dimension of the commutant in these cases.
Finally, we consider models with the $\pi$-bimagnon tower in the OBC one-dimensional spin-1 XY model as degenerate QMBS~\cite{schecter2019weak, mark2020eta, chandran2022review}, which resembles the $\eta$-pairing tower of states in the Hubbard model~\cite{yang1989eta, vafek2017entanglement, mark2020eta, moudgalya2020eta, pakrouski2020many}.
We consider the case where the commutant retains a $U(1)$ spin conservation symmetry, and in \cite{moudgalya2022from} we conjectured the local and commutant algebra pair
\begin{align}
    &\tmA^{(\XY)}_{\scar} = \lgen \{S^x_j S^x_{j+1} + S^y_j S^y_{j+1}\}, \{(S^z_j)^2\},\nn \\
    &\qquad\qquad\{(S^z_j + S^z_{j+1}) (1 - S^z_j S^z_{j+1})\} \rgen, \nn \\
    &\tmC^{(\XY)}_{\scar} = \lgen \{\ketbra{\Phi_m}{\Phi_n}\}, S^z_{\tot} \rgen,
\label{eq:spin1XYnondegpair}
\end{align}
where $S^{\alpha}_j$ are the spin-1 operators on site $j$, $\{\ket{\Phi_n}\}$ are the $L + 1$ QMBS states of the spin-1 XY model [see Eq.~(29) there for the definition], and $S^z_{\tot}$ is the total spin operator.
The dimension of $\tmC^{(\XY)}_{\scar}$ commutant can be obtained straightforwardly by counting the number of linearly independent operators, and we obtain $\dim(\tmC^{(\XY)}_{\scar}) = L(L+4)$.\footnote{There are $2L+1$ operators of the form $\{(S^z_{\tot})^m, m = 0, 1, \dots, 2L\}$ and $(L+1)^2$ operators of the form $\{\ketbra{\Phi_m}{\Phi_n}\}$, of which $2$ ($\ketbra{\Phi_0}$ and $\ketbra{\Phi_L}$) can be expressed as linear combinations of $\{(S^z_{\tot})^m\}$}.
In each of these cases, the commutant dimensions we expect are consistent with the results in Fig.~\ref{fig:liouvcomm}(c), and these also provide additional verifications to the commutants conjectured but not proven in \cite{moudgalya2022exhaustive}.
Note that in Fig.~\ref{fig:liouvcomm}(c), the system sizes studied in the spin-1 XY case are larger than for the AKLT tower of states since the bond algebra in the former case is generated by two-site terms as opposed to three-site terms.
\section{Inverse Problem: Constructing Symmetric Operators From the Symmetry}\label{sec:commutantloc}
For the sake of completeness, we discuss the inverse problem in the same language, i.e., given the generators of the commutant algebra $\mC$, the task is to construct local operators in the algebra $\mA$, which are by definition symmetric operators that possess the symmetry algebra $\mC$. 
As we discuss in Sec.~\ref{subsec:previousmethods}, this inverse method is completely equivalent to methods used in several previous works for the same purpose~\cite{obrien2016explicit, chertkov2018computational, qi2019determininglocal, yang2022detecting}.
In addition, there is a larger body of literature that focuses on recovering parent Hamiltonians for a general state, but their approaches are distinct from ours, e.g., they might assume additional structure of the state or resort to variational optimizations over some parameter space~\cite{greiter2018method, turkeshi2019entanglement, pakrouski2020automatic, bairey2019learning, dalmonte2018quantum, turkeshi2020parent}.
\subsection{Inverse Method}\label{subsec:inversemethod}
We first note that determining the space of \textit{all} symmetric operators is straightforward using the methods discussed in Secs.~\ref{sec:commutantblock} and \ref{sec:commutantLiouv}.
That is, the ``symmetry" (i.e., the Double Commutant Theorem) between $\mA$ and $\mC$ in Eqs.~(\ref{eq:Hilbertdecomp})-(\ref{eq:Cblock}) implies that the construction of the full algebra $\mA$ from the generators of $\mC$ can be done in exact analogy to the construction of $\mC$ from the generators of $\mA$.   
Focusing on the Liouvillian method, given a set of generators of the commutant algebra $\{\hQ_\alpha\}$, we can define superoperators as
\begin{equation}
    \hmL_{\hQ_\alpha} \defn \hQ_\alpha \otimes \mathds{1} - \mathds{1} \otimes \hQ_\alpha^T.
\label{eq:commsupop}
\end{equation}
Analogous to Sec.~\ref{sec:commutantLiouv}, the common kernel of $\{\hmL_{\hQ_\alpha}\}$, or equivalently, the ground states of the superoperator
\begin{equation}
\hmP' \defn \sum_{\alpha}{\hmL_{\hQ_\alpha}^\dagger \hmL_{\hQ_\alpha}}
\label{eq:Pprimdefn}
\end{equation}
are the linearly independent operators that span $\mA$.
However, for quantum matter applications, we are usually interested in constructing symmetric \textit{local} Hamiltonians, hence we wish to obtain the set of local operators in $\mA$.
Since any set of local operators with a finite range of at most $r_{\max}$ form a vector space, say $\mV_{\loc}$, this can be achieved using a direct analogy to the discussion in Sec.~\ref{subsec:conservedform}. 
In particular, we can compute $\hmP'_{\mV_{\loc}}$, the restriction of the matrix $\hmP'$ to $\mV_{\loc}$; this is defined analogous to Eq.~(\ref{eq:Prestmat}).
$\hmP'_{\mV_{\loc}}$ has properties analogous to Eq.~(\ref{eq:restrictedop}), i.e., its ground states are operators in the vector space $\mA \cap \mV_{\loc}$, which in turn are all the local operators in $\mA$ with range at most $r_{\max}$. 
\subsection{Connection to Previous Methods}\label{subsec:previousmethods}
We now elaborate on the precise relations of this inverse method to some of the methods introduced earlier in the literature for similar purposes.
First, in \cite{obrien2016explicit, chertkov2020engineering}, such symmetric local Hamiltonians were understood as zero modes of an appropriately constructed ``commutant matrix," which is precisely equivalent to the matrix $\hmP'_{\mV_{\loc}}$.
Second, \cite{qi2019determininglocal} introduced a ``correlation matrix" method for obtaining \textit{Hermitian} local operators that have a given state $\ket{\psi}$ as an eigenstate, which is equivalent to obtaining local operators that commute with $\ketbra{\psi}$. 
Applying our method to this problem, this is simply set of ``ground states" of the matrix $\hmP'_{\mV_{\loc}}$, where $\mV_{\loc} \defn \text{span}\{\oket{V_\mu}\}$ is a vector space spanned by local operators $\{V_\mu\}$ of interest. 
The matrix elements of $\hmP'_{\mV_{\loc}}$ are then given by
\begin{align}
    (\hmP'_{\mV_{\loc}})_{\mu, \mu'} &= \obra{V_{\mu}} \hmL^\dagger_{\ketbra{\psi}} \hmL_{\ketbra{\psi}} \oket{V_{\mu'}} \nn \\
       &= \text{Tr}\left([\ketbra{\psi}, V_{\mu}]^\dagger [\ketbra{\psi}, V_{\mu'}] \right)\nn \\
       &= \bra{\psi} \{V_\mu, V_{\mu'}\} \ket{\psi} - 2 \bra{\psi} V_\mu \ket{\psi} \bra{\psi} V_{\mu'} \ket{\psi}, 
\label{eq:correlationmatrix}
\end{align}
where $V_\mu$ are assumed to be Hermitian operators.
This is precisely the ``correlation matrix" defined in \cite{qi2019determininglocal} [see Eqs.~(1.1) and (2.7) there], and they too determine the local operators by determining the zero eigenvectors of the correlation matrix; hence these methods are completely equivalent when applied to determine Hermitian operators that have individual states $\ket{\psi}$ as eigenstates.
Finally, \cite{chertkov2018computational} independently introduced a ``covariance matrix" method where the ground states of the matrix are local operators that have a given state $\ket{\psi}$ as an eigenstate.
This covariance matrix differs from $\hmP'_{\mV_{\loc}}$ or the correlation matrix of Eq.~(\ref{eq:correlationmatrix}), and is defined as
\begin{equation}
    (\hmP''_{\mV_{\loc}})_{\mu,\mu'} \defn \bra{\psi} V_\mu V_{\mu'} \ket{\psi} - \bra{\psi} V_{\mu} \ket{\psi} \bra{\psi} V_{\mu'} \ket{\psi},
\label{eq:covariancematrix}
\end{equation}
where $\{V_\mu\}$ is a linearly independent set of Hermitian operators that spans $\mV_{\loc}$.
Any ground state of $(\hmP''_{\mV_{\loc}})$ (i.e., with eigenvalue $0$) corresponds to an operator in $\mV_{\loc}$ that has $\ket{\psi}$ as an eigenstate.
While such an operator might be non-Hermitian in general, Hermitian operators can be obtained by restricting to \textit{real} eigenvectors in the ground state space.
Using the fact that $\hmP'_{\mV_{\loc}} = \hmP''_{\mV_{\loc}} + (\hmP''_{\mV_{\loc}})^\ast$ and that $\hmP''_{\mV_{\loc}}$ and $\hmP'_{\mV_{\loc}}$ are both positive-semidefinite Hermitian matrices, it is easy to show that the subspace spanned by all the ground states of $\hmP'_{\mV_{\loc}}$ exactly coincides with the subspace spanned by real ground states of $\hmP''_{\mV_{\loc}}$.\footnote{A quick proof is as follows.
We wish to show $(\hmP'_{\mV_{\loc}})_{\mu,\nu} x_{\nu} = 0 \iff (\hmP''_{\mV_{\loc}})_{\mu,\nu} x_\nu = 0$, where repeated indices are summed over and $x_\nu$ is assumed to be real in the R.H.S.
Starting with the L.H.S., the relation between $\hmP''_{\mV_{\loc}}$ and $\hmP'_{\mV_{\loc}}$ shows that $(\hmP''_{\mV_{\loc}})_{\mu, \nu} x_{\nu} = - (\hmP''_{\mV_{\loc}})^\ast_{\mu,\nu} x_{\nu} \implies x_{\mu} (\hmP''_{\mV_{\loc}})_{\mu, \nu} x_{\nu} = - x_{\mu}(\hmP''_{\mV_{\loc}})^\ast_{\mu,\nu} x_{\nu} \implies x_{\mu} (\hmP''_{\mV_{\loc}})_{\mu, \nu} x_{\nu} = 0$, where we have used the Hermiticity of $\hmP''_{\mV_{\loc}}$. 
Noting that $\hmP'_{\mV_{\loc}}$ is a real symmetric matrix, we have $x_{\mu} = x_{\mu}^\ast$ (all its eigenspaces can be chosen to be real), hence $x_{\mu}^\ast (\hmP''_{\mV_{\loc}})_{\mu,\nu} x_\nu = 0$, which is equivalent to the R.H.S. since $\hmP''_{\mV_{\loc}}$ is p.s.d.
For the other direction, we start with the R.H.S.\ and assume $x_{\mu} = x_{\mu}^\ast$ to obtain. $(\hmP''_{\mV_{\loc}})^\ast_{\mu,\nu} x_{\nu} = 0$.
Using the expression for $\hmP'_{\mV_{\loc}}$, the L.H.S.\ follows.}
Hence, if we are interested in Hermitian local operators, and we interpret the procedure of \cite{chertkov2018computational} as solving for the real eigenvectors, these methods are equivalent.
When applied to states with an MPS representation, these methods are also closely related to the tensor network method discussed in~\cite{yang2022detecting}, which determines strictly local or extensive local (emergent) symmetry operators given the MPS ground state of a certain Hamiltonian.
\subsection{Construction of Type I and Type II Symmetric Operators}
We now discuss how to apply this method to distinguish two types of symmetric operators that can be constructed out of a \textit{bond algebra} $\mA \defn \lgen \{\hH_\alpha\} \rgen$,  i.e.,  an algebra generated by a set of \textit{strictly local} operators $\{\hH_\alpha\}$, where none of the $\hH_\alpha$ are extensive local. 
To recap, in \cite{moudgalya2022exhaustive} we found qualitatively new types of symmetric Hamiltonians (i.e., extensive local operators) corresponding to symmetry algebras that are ``unconventional",  for example, commutant algebras that explain QMBS. 
One obvious class of symmetric extensive local operators are those that can be expressed as a sum of symmetric strictly local operators, and we refer to these as \textit{Type I} symmetric Hamiltonians. 
\textit{Type II} symmetric Hamiltonians are then those that cannot be expressed as a sum of symmetric strictly local operators, i.e., they necessarily involve highly non-local expressions in terms of the strictly local generators of $\mA$.
In a previous work~\cite{moudgalya2022from},  we showed that for commutants generated by on-site unitary operators,  all symmetric Hamiltonians are of type I,  whereas in another previous work~\cite{moudgalya2022exhaustive} we showed in the case of QMBS that there are Hamiltonians that are type II. 
Further, we introduced the notion of \textit{equivalence classes} of type II symmetric operators, where two type II operators are equivalent if they differ by the addition of a type I symmetric operator.
Since type I symmetric operators of a given range form a vector space that is a subspace of the vector space of all symmetric operators of that range, the set of equivalence classes of type II operators has an appropriate quotient space structure.
These equivalence classes of type II operators of range at most $r_{\max}$ can be directly extracted numerically using the inverse methods discussed in this section, as we now discuss.
Note that similar ideas were used to discover various Hamiltonians with various examples of QMBS in \cite{mark2020eta, odea2020from, ren2021deformed}. 
Concretely,  we start by considering the set of all clusters of sites on the lattice of range $r_{\max}$, which we denote by $\{R\}$,  and the associated vector spaces $\{\mV_{R}\}$ of strictly local operators with support strictly within the respective clusters.
We can then apply the method of Sec.~\ref{subsec:inversemethod} using $\mV_{\loc} = \sum_R{\mV_R}$ to compute the vector space $\mO^{(\mA)}_{\loc}$ of \textit{all} operators in $\mA$ that are linear combinations of strictly local operators with support on any one of the clusters $\{R\}$ (operators with different supports among $\{R\}$ can appear in the sum).
Note that $\mO^{(\mA)}_{\loc}$ is the space of all symmetric local operators, both strictly local and extensive local, of range at most $r_{\max}$, and it includes both type I and type II symmetric operators.
(Here and below, the dependence of the discussed operator spaces on $r_{\max}$ is understood implicitly.)
To separate the type I and type II operators, we can then apply the method of Sec.~\ref{subsec:inversemethod} restricting to each such cluster $R$, i.e., by using $\mV_{\loc} = \mV_R$.
This yields the vector space of operators in $\mA$ that have support only on the cluster $R$, which we denote by $\mO^{(\mA)}_R$. 
By construction,  the vector space $\mO^{(\mA)}_{\text{I}} \defn \sum_R{\mO^{(\mA)}_R}$ is the vector space of all type I symmetric operators of range at most $r_{\max}$, which includes both strictly local and extensive local operators. 
With these vector spaces $\mO^{(\mA)}$ and $\mO^{(\mA)}_{\text{I}}$, we directly obtain the equivalence classes of type II symmetric operators, i.e., the quotient space $\mO^{(\mA)}/\mO^{(\mA)}_{\text{I}}$, whose dimension is given by the difference $N_{\text{II}} \defn \dim(\mO^{(\mA)}/\mO^{(\mA)}_{\text{I}}) = \dim(\mO^{(\mA)}) - \dim(\mO^{(\mA)}_{\text{I}})$.
Numerically applying the procedure to standard examples of on-site unitary symmetries, we then recover that $N_{\text{II}} = 0$ for all choices $r_{\max} \ll L$, consistent with the proof that there are no Type II symmetric operators in such cases~\cite{moudgalya2022from}.
On the other hand, applying this method to unconventional symmetries such as QMBS, we find in certain cases that $N_{\text{II}}$ can increase with $r_{\max}$, which shows that several independent type II symmetric operators can exist, as discussed in \cite{moudgalya2022exhaustive}.
\section{Conclusions and Outlook}\label{sec:conclusions}
In this work, we provided two methods to numerically construct commutant algebras corresponding to \textit{families} of Hamiltonians. 
One of these involves simultaneous block diagonalization of two randomly chosen operators in the family and builds on earlier works of similar nature~\cite{chertkov2020engineering, nie2021operator}.
The other method maps this onto a problem of determining the frustration-free ground state of a Liouvillian superoperator, which can be efficiently solved in one dimension using MPS-based techniques discussed in \cite{yao2022bounding}.
These methods are useful in determining all the symmetries or dynamically disconnected ``Krylov subspaces" of a particular family of systems, and we demonstrate this by applying these methods to several examples where we detect the presence of conventional symmetries~\cite{moudgalya2022from}, Hilbert space fragmentation~\cite{moudgalya2021hilbert}, or Quantum Many-Body Scars~\cite{moudgalya2022exhaustive}. 
In addition, they allow us to conjecture and corroborate commutants corresponding to local algebras we study in cases where we are not able to provide a proof, and could also be useful in other contexts, e.g., in quickly checking if a family of Hamiltonians has some unexpected symmetries.
Finally, we also discussed inverse methods to determine local symmetric operators given the set of generators of a symmetry algebra. 
These can be useful in determining the exhaustive set of generators for a local algebra corresponding to a given symmetry algebra, or for identifying distinct types of local symmetric operators.
While in this work we have adapted a ``proof-of-principle" approach to demonstrate the numerical methods, we believe there are many avenues to make these methods much more efficient and hence more widely applicable to physically relevant families of Hamiltonians.
Moreover, as evident from some of the examples, even the efficient Liouvillian method  of Sec.~\ref{subsec:1deff} in one dimension works best in practice only when the generators of the local algebra are strictly local and with open boundary conditions.
The addition of terms with periodic boundary conditions or extensive local terms is naively a significant hindrance, and it would be interesting to explore tricks that might make those cases  computationally more tractable.
It would also be interesting to generalize this method to two dimensions, where many  Hamiltonians of physical interest lie, and perhaps ideas from the theory of Projected Entangled Pair States (PEPS)~\cite{schuch2010peps} might be useful.
Numerical methods for determining the commutant are also useful for scanning through physically relevant models looking for ``unconventional symmetries," which includes scars and fragmentation.
Indeed, in an upcoming work~\cite{szminprep}, we apply such methods to discover examples of Strong Zero Modes (SZM)~\cite{alicea2016topological, fendley2016strong} that can be understood within the commutant algebra framework.
This in turn allows us to construct \textit{non-integrable} models with SZM, settling the debate of whether SZM can only occur in integrable models.
There are multiple questions on the theoretical front too.
First, it would be interesting to connect the  methods discussed here to other methods introduced in the literature for identifying unconventional symmetries such as quantum many-body scars and Hilbert space fragmentation.
For example, \cite{regnault2022integer} used integer factorizations of characteristic polynomials of Hamiltonians to detect Hilbert space fragmentation and quantum many-body scars, while \cite{szoldra2022unsupervised} used machine learning methods to detect quantum many-body scars. 
Furthermore, the Liouvillian method here shows that symmetry algebras can be understood as frustration-free ground state manifolds of local superoperators.
This suggests that this problem might be analytically tractable, and indeed we find examples of such cases, which leads to several insights on symmetric systems with locality, and we will report these results elsewhere~\cite{moudgalya2023inprep}.
Also, given that several types of conventional and unconventional symmetries can be understood within the commutant algebra framework~\cite{moudgalya2021hilbert, moudgalya2022from, moudgalya2022exhaustive}, it is natural to wonder if the frustration-free ground state property introduces some general constrains on the kind of operators that are allowed to be symmetries, e.g., do symmetry operators necessarily have low operator entanglement?
We defer explorations of such questions to future work.
\section*{Acknowledgements}
We particularly thank Pengfei Zhang for useful discussions and for sharing unpublished notes on \cite{yao2022bounding}.
We also thank Bryan Clark, Nick O'Dea, Laimei Nie, and Frank Pollmann for useful discussions.
This work was supported by the Walter Burke Institute for Theoretical Physics at Caltech; the Institute for Quantum Information and Matter, an NSF Physics Frontiers Center (NSF Grant PHY-1733907); and the National Science Foundation through grant DMR-2001186.
S.M. acknowledges the hospitality of the Centro de Ciencias de Benasque Pedro Pascual, where a part of this work was completed. 
\bibliography{newrefs}
\appendix 
\onecolumngrid
\section{Constructing Non-Abelian Commutants Using Block-Diagonalization}\label{app:nonabelianblock}
In this appendix, we give details on constructing non-Abelian commutants using the simultaneous block diagonalization method discussed in Sec.~\ref{sec:commutantblock}. 
According to Eq.~(\ref{eq:Ablock}) the Hamiltonians $H^{(1)}$ and $H^{(2)}$ can be unitarily transformed into a basis where they are simultaneously block-diagonal, i.e.,
\begin{equation}
    W^\dagger H^{(1)} W = \bigoplus_\lambda [M^{(1)}_{D_\lambda} \otimes \mathds{1}_{d_\lambda}],\;\;\; W^\dagger H^{(2)} W = \bigoplus_\lambda [M^{(2)}_{D_\lambda} \otimes \mathds{1}_{d_\lambda}],\;\;\; W^\dagger W = \mathds{1}.
\label{eq:H1H2forms}
\end{equation}
In the following, we assume that the blocks labelled by $\lambda$ have been ``resolved" using methods discussed in Sec.~\ref{subsec:Zextract}, and that the numbers $d_\lambda$ and $D_\lambda$ have already been extracted using the methods described in Sec.~\ref{subsec:Ddextract}.
The tensor product form used to write the expected finer block-diagonal structure inside the block $\lambda$ in the R.H.S.'s of the above equations assumes a formal factoring of the corresponding $D_\lambda d_\lambda$-dimensional space as $\mH_\lambda^{(\mA)} \otimes \mH_\lambda^{(\mC)}$, where the formal spaces $\mH_\lambda^{(\mA)}$ and $\mH_\lambda^{(\mC)}$ have dimensions $D_\lambda$ and $d_\lambda$ respectively;
$M_{D_\lambda}^{(1)}$ and $M_{D_\lambda}^{(2)}$ are operators in $\mH_\lambda^{(\mA)}$, while $\mathds{1}_{d_\lambda}$ is the identity in $\mH_\lambda^{(\mC)}$.
However, this tensor factoring of the space is not initially known to us other than that it exists.
According to Eq.~(\ref{eq:Cblock}), the unitarily-transformed commutant $W^\dagger \mC W$ restricted to the block labelled by $\lambda$ is spanned by operators of the form $\mathds{1}_{D_\lambda} \otimes N_{d_\lambda}$ where $N_{d_\lambda}$ is an arbitrary operator in $\mH_\lambda^{(\mC)}$.
Suppose $\{\ket{v_\beta}\}$, $1 \leq \beta \leq d_\lambda$, is a complete orthonormal basis in this space.
Denoting this restricted commutant in the original computational basis by $\mC_{\lambda}$, it can be written as
\begin{equation}
    \mC_{\lambda} = \text{span}_{\beta,\beta'} \{ W (\mathds{1} \otimes \ketbra{v_\beta}{v_{\beta'}} ) W^\dagger \} 
    = \text{span}_{\beta,\beta'} \{ W 
    (\mathds{1} \otimes U_{d_\lambda})
    (\mathds{1} \otimes \ketbra{v_\beta}{v_{\beta'}} )
    (\mathds{1} \otimes U_{d_\lambda}^\dagger)
    W^\dagger \}
    ~,
    \label{eq:Clambdaspan}
\end{equation}
where in the last form $U_{d_\lambda}$ can be any fixed unitary acting in $\mH_\lambda^{(\mC)}$.
While these operators in $\mC_\lambda$ can be directly constructed if the $W$ or the block-diagonal basis is known, we do not have a direct access to such information.
What we have instead is how the instances $H^{(1)}$ and $H^{(2)}$ act in our computational basis, in particular we can diagonalize $H^{(1)}$ and evaluate expectations values of $H^{(2)}$ in the eigenstates of $H^{(1)}$.
In particular, diagonalizing $H^{(1)}$ of the form of Eq.~(\ref{eq:H1H2forms}) and assuming there are no degeneracies in the spectrum of $M^{(1)}_{D_\lambda}$, we obtain its eigenstates and $d_\lambda$-fold degenerate eigenvalues $\{m_\alpha\}$:
\begin{equation}
    H^{(1)}\sket{\phi_{\alpha\beta}} = m_\alpha\sket{\phi_{\alpha\beta}},\;\;\;1 \leq \alpha \leq D_\lambda,\;\;1 \leq \beta \leq d_\lambda,\;\;\sket{\phi_{\alpha\beta}} = W(\sket{x_\alpha} \otimes \sket{y^{(\alpha)}_\beta}),
\label{eq:H1diag}
\end{equation}
where $\{m_\alpha\}$ and $\{\ket{x_\alpha}\}$ are the eigenvalues and eigenvectors of $M^{(1)}_{D_\lambda}$, and for each $\alpha$ we have $\textrm{span}_\beta\{\sket{y^{(\alpha)}_\beta}\} =  \text{span}_\beta\{\ket{v_\beta}\}$.
Note that in Eq.~(\ref{eq:H1diag}), while we know the vectors $\{\ket{\phi_{\alpha\beta}}\}$, we do not yet know the unitary $W$ or the factorization of the transformed states $W^\dagger \ket{\phi_{\alpha\beta}}$, but we are guaranteed that it exists. Also, $\{\sket{y_\beta^{(\alpha)}} \}$ can be an arbitrary orthonormal basis in $\mH_\lambda^{(\mC)}$ since diagonalizing $H^{(1)}$ only gives us an arbitrary basis in the degenerate space corresponding to its eigenvalue $m_\alpha$.
Our task is to construct $\mC_\lambda$ using only the information we have, which we now develop.
Since $\sket{y^{(\alpha)}_\beta} = U^{(\alpha)}\ket{v_\beta}$ for some unitary $U^{(\alpha)}$, we can express
\begin{equation}
    \mathds{1} \otimes \ketbra{v_\beta}{v_{\beta'}} = \sumal{\alpha}{}{[\mathds{1} \otimes (U^{(\alpha)})^\dagger] W^\dagger \ketbra{\phi_{\alpha\beta}}{\phi_{\alpha\beta'}}W[\mathds{1} \otimes U^{(\alpha)}] },
\label{eq:Piphirelation}
\end{equation}
where we have used that $\{\ket{x_\alpha}\}$ form a complete orthonormal basis in $\mH_\lambda^{(\mA)}$.
Choosing the fixed unitary $U_{d_\lambda}$ in Eq.~(\ref{eq:Clambdaspan}) to be $U^{(\alpha_0)}$ with some fixed $\alpha_0$, the commutant $\mC_\lambda$ is spanned by
\begin{gather}
    \tPi_{\beta,\beta'} \defn W [\mathds{1} \otimes U^{(\alpha_0)}] (\mathds{1} \otimes \ketbra{v_\beta}{v_{\beta'}}) [\mathds{1} \otimes (U^{(\alpha_0)})^\dagger]W^\dagger = \sumal{\alpha}{}{\tU^{(\alpha_0\alpha)} \ketbra{\phi_{\alpha\beta}}{\phi_{\alpha\beta'}} (\tU^{(\alpha_0\alpha)})^\dagger},\;\;
    \label{eq:Pitilddefn}\\
    \tU^{(\alpha_0\alpha)} \defn W[\mathds{1}\otimes U^{(\alpha_0)} (U^{(\alpha)})^\dagger]W^\dagger.
\label{eq:Utilddefn}
\end{gather}
Inserting decompositions of identity in terms of these states in Eq.~(\ref{eq:Pitilddefn}), we obtain
\begin{equation}
    \tPi_{\beta,\beta'} = \sumal{\alpha, \gamma, \delta, \gamma',\delta'}{}{\ketbra{\phi_{\gamma\delta}}{\phi_{\gamma\delta}}\tU^{(\alpha_0\alpha)} \ketbra{\phi_{\alpha\beta}}{\phi_{\alpha\beta'}} (\tU^{(\alpha_0\alpha)})^\dagger\ketbra{\phi_{\gamma'\delta'}}{\phi_{\gamma'\delta'}}}.
\label{eq:Pitildinsert}
\end{equation}
Using Eqs.~(\ref{eq:Utilddefn}) and (\ref{eq:H1diag}), we obtain
\begin{gather}
    \bra{\phi_{\gamma\delta}}\tU^{(\alpha_0\alpha)}\ket{\phi_{\alpha\beta}} = \delta_{\gamma\alpha} \sbra{y^{(\gamma)}_\delta} U^{(\alpha_0)}(U^{(\alpha)})^\dagger\sket{y^{(\alpha)}_\beta} = \delta_{\gamma\alpha} \sbra{v_\delta} (U^{(\alpha)})^\dagger U^{(\alpha_0)}\sket{v_\beta} \defn \delta_{\gamma\alpha} V^{(\alpha\alpha_0)}_{\delta\beta} \nn\\
    \bra{\phi_{\alpha\beta'}}(\tU^{(\alpha_0\alpha)})^\dagger\ket{\phi_{\gamma'\delta'}} = \delta_{\alpha\gamma'} \sbra{y^{(\alpha)}_{\beta'}} U^{(\alpha)} (U^{(\alpha_0)})^\dagger\sket{y^{(\gamma')}_{\delta'}} = \delta_{\alpha\gamma'} \sbra{v_{\beta'}} (U^{(\alpha_0)})^\dagger U^{(\alpha)} \sket{v_{\delta'}} = \delta_{\alpha\gamma'} V^{(\alpha_0\alpha)}_{\beta'\delta'},
\label{eq:Vtildmatrixels}
\end{gather}
where we have defined $V^{(\alpha'\alpha)}_{\beta'\beta} \defn \bra{v_{\beta'}} V^{(\alpha'\alpha)} \ket{v_{\beta}}$ and $V^{(\alpha'\alpha)} \defn (U^{(\alpha')})^\dagger U^{(\alpha)}$, a $d_\lambda$-dimensional unitary matrix.
Substituting Eq.~(\ref{eq:Vtildmatrixels}) in (\ref{eq:Pitildinsert}), we obtain
\begin{equation}
    \tPi_{\beta,\beta'} = \sumal{\alpha, \delta, \delta'}{}{\ket{\phi_{\alpha\delta}}V^{(\alpha\alpha_0)}_{\delta\beta}V^{(\alpha_0\alpha)}_{\beta'\delta'}\bra{\phi_{\alpha\delta'}}} = \sumal{\alpha, \delta, \delta'}{}{(V^{(\alpha_0\alpha)}_{\beta\delta})^\ast V^{(\alpha_0\alpha)}_{\beta'\delta'}\ketbra{\phi_{\alpha\delta}}{\phi_{\alpha\delta'}}}.
\label{eq:PitildviaVV}
\end{equation}
We now show that we can relate $\{V^{(\alpha'\alpha)}_{\beta'\beta}\}$ to the matrix elements of $H^{(2)}$ in the $\{\ket{\phi_{\alpha\beta}}\}$ basis, i.e., the eigenbasis of $H^{(1)}$.
As a consequence of Eqs.~(\ref{eq:H1H2forms}) and (\ref{eq:H1diag}), the matrix elements of $H^{(2)}$ read
\begin{equation}
    \bra{\phi_{\alpha'\beta'}}H^{(2)}\ket{\phi_{\alpha\beta}} = \bra{x_{\alpha'}} M^{(2)}_{D_\lambda} \ket{x_\alpha} \sbraket{y^{(\alpha')}_{\beta'}}{y^{(\alpha)}_{\beta}} = \bra{x_{\alpha'}} M^{(2)}_{D_\lambda} \ket{x_\alpha} \bra{v_{\beta'}} (U^{(\alpha')})^\dagger U^{(\alpha)} \ket{v_{\beta}} = \bra{x_{\alpha'}} M^{(2)}_{D_\lambda} \ket{x_\alpha} V^{(\alpha'\alpha)}_{\beta'\beta}.
\label{eq:H2matrix}
\end{equation}
The ratios of the matrix elements of $V^{(\alpha'\alpha)}$ can then be written in terms of matrix elements of $H^{(2)}$, which completely determines the matrix $V^{(\alpha'\alpha)}$ up to a single non-zero element.
That is,
\begin{equation}
    V^{(\alpha'\alpha)} =  c_{\alpha'\alpha} G^{(\alpha'\alpha)},\;\;\; G^{(\alpha'\alpha)}_{\beta'\beta} \defn \bra{\phi_{\alpha' \beta'}} H^{(2)}\ket{\phi_{\alpha\beta}},\;\;c_{\alpha'\alpha} \defn [\bra{x_{\alpha'}} M^{(2)}_{D_\lambda} \ket{x_\alpha}]^{\text{-}1}.
\label{eq:Vmat}
\end{equation}
The absolute value of $c_{\alpha'\alpha}$ can be obtained by imposing the unitarity of $V^{(\alpha'\alpha)}$, 
\begin{equation}
(G^{(\alpha'\alpha)})^\dagger G^{(\alpha'\alpha)} = \frac{1}{|c_{\alpha'\alpha}|^2 }\mathds{1}\;\;\implies\;\;|c_{\alpha'\alpha}|^2 = \big|\det G^{(\alpha'\alpha)}\big|^{-\frac{2}{d_\lambda}} ~.
\end{equation}
Plugging this into Eq.~(\ref{eq:Vmat}), the matrix elements $V^{(\alpha'\alpha)}_{\beta'\beta}$ can be expressed as
\begin{equation}
    V^{(\alpha'\alpha)}_{\beta'\beta} = \frac{1}{\big|\det G^{(\alpha'\alpha)}\big|^{\frac{1}{d_\lambda}}}\bra{\phi_{\alpha' \beta'}} H^{(2)}\ket{\phi_{\alpha\beta}},
\label{eq:Vmatfinal}
\end{equation}
where we have ignored the phase factor, which can be arbitrary and does not enter into the expression for the operators in the commutant.
We can then rewrite Eq.~(\ref{eq:PitildviaVV}) as 
\begin{equation}
\tPi_{\beta,\beta'} = \sumal{\alpha, \delta, \delta'}{}{|c_{\alpha_0\alpha}|^2 (G^{(\alpha_0\alpha)}_{\beta\delta})^\ast G^{(\alpha_0\alpha)}_{\beta'\delta'}\ketbra{\phi_{\alpha\delta}}{\phi_{\alpha\delta'}}} = \sumal{\alpha, \delta, \delta'}{}{\big|\det G^{(\alpha_0\alpha)}\big|^{-\frac{2}{d_\lambda}} (G^{(\alpha_0\alpha)}_{\beta\delta})^\ast G^{(\alpha_0\alpha)}_{\beta'\delta'}\ketbra{\phi_{\alpha\delta}}{\phi_{\alpha\delta'}}},
\label{eq:Pitildfinal}
\end{equation}
which are all in terms of ``known" quantities. 
This allows us to construct $ \mC_\lambda = \textrm{span}_{\beta,\beta'}\{\tPi_{\beta,\beta'}\}$.
Repeating this procedure for all the blocks labelled by different $\lambda$'s allows us to construct the full commutant $\mC$.
\section{Details on Efficient Construction of Commutants in One Dimension}\label{app:efficientcommutant1d}
In this appendix, we provide some details on an efficient Liouvillian method to construct the commutant algebra in one-dimensional systems, discussed in Sec.~\ref{sec:commutantLiouv}. 
This method works best for commutants of bond algebras generated by strictly local terms with OBC, which we discuss in Apps.~\ref{subsec:2bondalgebracomm} and \ref{subsec:nbondalgebracomm}. 
In Apps.~\ref{subsec:PBCbondalgebra} and \ref{subsec:localalgebra} we discuss the extensions of this method to algebras generated by PBC terms and to local algebras where some of the generators are extensive local operators.
\subsection{Bond algebras generated by nearest-neighbor terms with OBC}\label{subsec:2bondalgebracomm}
We first illustrate this method for bond algebras generated by nearest-neighbor terms with OBC.
In this case, the algebra generators $\{\hH_\alpha\}$ and the superoperators $\{\hmP_{\hH_\alpha}\}$ discussed in Sec.~\ref{sec:commutantLiouv} are strictly local nearest-neighbor terms, which we denote by $\{\hH_{j,j+1}\}_{j = 1}^{L-1}$ and $\{\hmP_{j,j+1}\}_{j = 1}^{L-1}$, where $L$ is the system size. 
Our aim is to construct the commutant recursively, i.e., to obtain the commutant of an $N$-site system from the commutant of an $(N-1)$-site system. 
In the following, we denote the commutant of a system of size $n$ as
\begin{equation}
  \mC_n \defn \text{span}\{\oket{C^{(n)}_{\mu_n}}\}_{\mu_n = 1}^{\chi_n},\;\;\chi_n \defn \dim(\mC_n),\;\;\obraket{C^{(n)}_{\alpha}}{C^{(n)}_{\beta}} = \delta_{\alpha,\beta},
\label{eq:CNdefnnorm}
\end{equation}
where $\obraket{O_1}{O_2}\defn \frac{1}{D}{\text{Tr}(O_1^\dagger O_2)}$ is the usual Hilbert-Schmidt overlap of two operators, and $D$ is the dimension of the Hilbert space in which the operators $O_1$ and $O_2$ act.
Note that $\{\oket{C^{(n)}_{\mu_n}}\}$ are operators on the Hilbert space of $n$ sites, hence they are linear combinations of computational basis operators $\oket{m_1 \cdots m_n}$, $1 \leq m_j \leq d^2_{\loc}$, where $d_{\loc}$ is the on-site Hilbert space dimension assumed for simplicity to be the same on all sites.
To construct the commutant recursively, it is convenient to think of $\{\oket{C^{(N)}_{\mu_N}}\}$ as a Matrix Product State (MPS), see Fig.~\ref{fig:MPScomm}a.
In principle, an MPS can be understood via successive Schmidt decompositions of the ``vector" $\oket{C^{(N)}_{\mu_N}}$.
To begin, the decomposition w.r.t. a bipartition of the chain into the left $(N-1)$ sites and the rightmost $N$-th site is of the form $\oket{C^{(N)}_{\mu_N}} = \sum_\alpha{\oket{L^\alpha_{1, \cdots, N-1}} \otimes \oket{R^\alpha_N}}$, where $\oket{L^\alpha_{1, \cdots, N-1}}$ and $\oket{R^\alpha_N}$ are Schmidt vectors with supports on the left and right partitions respectively.
Since $\{\hmP_{j, j+1}\}_{j = 1}^{N-1}$ vanish on $\oket{C^{(N)}_{\mu_N}}$ by definition, they must also vanish on the corresponding left Schmidt vectors $\{\oket{L^\alpha_{1, \cdots, N-1}}\}$.
Hence the left Schmidt vectors $\{\oket{L^\alpha_{1, \cdots, N-1}}\}$ are in the commutant $\mC_{N-1}$ and can be expressed as linear combinations of $\oket{C^{(N-1)}_{\mu_{N-1}}}$, and the right Schmidt vectors $\{\oket{R^\alpha_N}\}$ are a linear combinations of the computational basis vectors $\{\oket{m_j}\}$.
With this in mind, $\oket{C^{(N)}_{\mu_N}}$ can always be expressed in terms of $\{\oket{C^{(N-1)}_{\mu_{N-1}}}\}$ as
\begin{equation}
    \oket{C^{(N)}_{\mu_N}} = \sumal{m_N = 1}{d^2_{\loc}}{\sumal{\mu_{N-1} = 1}{\chi_{N-1}}{\oket{C^{(N-1)}_{\mu_{N-1}}} \otimes \opket{m_N} [A^{[m_N]}_N]_{\mu_{N-1}\mu_N}}},
\label{eq:CNrecursive}
\end{equation}
where $[A^{[m_N]}_{N}]_{\mu_{N-1}\mu_N}$ can be viewed as elements of some tensor $A_N$ with a $d^2_{\loc}$-dimensional physical index labelled by $m_N$, and two auxiliary indices of dimensions $\chi_{N-1}$ and $\chi_N$ labelled by $\mu_{N-1}$ and $\mu_N$ respectively.
We can then repeatedly apply Eq.~(\ref{eq:CNrecursive}), e.g., applying twice we obtain
\begin{equation}
    \oket{C^{(N)}_{\mu_N}} = \sumal{\mu_{N-2}, \mu_{N-1}}{}{\sumal{m_{N-1}, m_N}{}{\oket{C^{(N-2)}_{\mu_{N-2}}} \otimes \oket{m_{N-1} m_{N}}\ [A^{[m_{N-1}]}_{N-1}]_{\mu_{N-2}\mu_{N-1}}[A^{[m_{N}]}_{N}]_{\mu_{N-1}\mu_{N}}}},
\label{eq:CNtwice}
\end{equation}
and applying $N$ times, we obtain the MPS form of $\oket{C^{(N)}_{\mu_N}}$:
\begin{equation}
\oket{C^{(N)}_{\mu_N}} = \sumal{\{m_j\}_{j = 1}^N, \{\mu_j\}_{j = 1}^{N-1}}{}{[A^{[m_1]}_1]_{\mu_1} [A^{[m_2]}_{2}]_{\mu_1\mu_2} [A^{[m_3]}_3]_{\mu_2\mu_3} \cdots [A^{[m_N]}_N]_{\mu_{N-1}\mu_N} \oket{m_1 m_2 \cdots m_N}},
\label{eq:CNMPSform}
\end{equation}
which we show pictorially in Fig.~\ref{fig:MPScomm}(a).
Note that it is convenient to view $\{A^{[m_j]}_j\}_{j = 2}^N$ as a $\chi_{j-1} \times \chi_j$ matrix over the auxiliary indices, and the leftmost tensor $A^{[m_1]}_1$ as a $\chi_1$-dimensional vector with a single auxiliary index, although we will sometimes implicitly assign a dummy auxiliary index to $A_1$ and treat it as a $\chi_0 \times \chi_1$ matrix where $\chi_0 \defn 1$.
To construct the commutant $\mC_N$, we hence need to solve for the tensors $\{A_j\}_{j = 1}^N$. 
We start with $N = 2$, and directly solve for the tensors $A_1$ and $A_2$ as follows. 
The vectors $\{\oket{C^{(2)}_{\mu_2}}\}$ are defined as the orthonormal span of the kernel of $\hmP_{1,2}$, which can be obtained by a direct diagonalization. 
To construct the individual tensors, we can perform a Schmidt decomposition similar to Eq.~(\ref{eq:CNrecursive}) on the vectors $\oket{C^{(2)}_{\mu_2}}$ to obtain
\begin{equation}
    \oket{C^{(2)}_{\mu_2}} = \sumal{\mu_1,m_2}{}{[A^{[m_2]}_2]_{\mu_1\mu_2} \oket{C^{(1)}_{\mu_1}} \otimes \oket{m_2}},\;\;\;\oket{C^{(1)}_{\mu_1}} = \sum_{m_1 = 1}^{d^2_{\loc}}{[A^{[m_1]}_{1}]_{\mu_1}\oket{m_1}},
\label{eq:twositeschmidt}
\end{equation}
where $\oket{C^{(1)}_{\mu_1}}$ is a vector (which does not have any interpretation as a basis vector of a commutant) with support only on the first site.
Note that in Eq.~(\ref{eq:twositeschmidt}), the simultaneous Schmidt decomposition of $\{\oket{C^{(2)}_{\mu_2}}\}$ (achieved using the Singular Value Decomposition of properly reshaped amplitudes lumping in the index $\mu_2$) gives us the tensor $A_2$ and the orthonormal vectors $\{\oket{C^{(1)}_{\mu_1}}\}$, and the expressions of $\{\oket{C^{(1)}_{\mu_1}}\}$ in the computational basis gives us the tensor $A_1$.
The condition that $\hmP_{1,2}$ satisfies is shown pictorially in Fig.~\ref{fig:MPScomm}(b).
Given the initial tensors, we can solve for the remaining tensors recursively, using ideas similar to those used in Density Matrix Renormalization Group (DMRG) algorithms~\cite{schollwock2011density}.
Suppose we know the operators $\oket{C^{(N-1)}_{\mu_{N-1}}}$ in the $(N-1)$-site commutant $\mC_{N-1}$, i.e., we have the tensors $\{A_j\}_{j=1}^{N-1}$. 
We can solve for the tensor $A_N$ by requiring 
\begin{equation}
    \hmP_{N-1, N}\oket{C^{(N)}_{\alpha}} = 0\;\;\iff\;\;\obra{C^{(N)}_{\alpha'}} \hmP_{N-1, N} \oket{C^{(N)}_{\alpha}} = 0,\quad \forall \alpha,\alpha'.
\label{eq:CNreqdcondition}
\end{equation}
Note that the implication uses the positive semi-definite property of $\hmP_{N-1,N}$, where it is easy to show that $\obra{C^{(N)}_\alpha}\hmP_{N-1,N}\oket{C^{(N)}_\alpha} \implies \hmP_{N-1, N}\oket{C^{(N)}_\alpha} = 0$.
Moreover, since $\hmP_{N-1, N}$ is a two-site operator, it is convenient to express $\{\oket{\mC^{(N)}_{\mu_N}}\}$ as in Eq.~(\ref{eq:CNtwice}), in which case Eq.~(\ref{eq:CNreqdcondition}) can be written as (relabelling $\alpha,\alpha' \to \mu_N,\mu_N'$)
\begin{equation}
    \sumal{\substack{\mu\\
    \mu'_{N-1}, \mu_{N-1}}}{}{\sumal{\substack{m_{N-1}, m_N\\
    m'_{N-1}, m'_N}}{}{\obra{m'_{N-1} m'_N} \hmP_{N-1, N} \oket{m_{N-1} m_N} [A^{[m'_{N-1}]}_{N-1}]^\ast_{\mu \mu'_{N-1}} [A^{[m'_{N}]}_N]^\ast_{\mu'_{N-1}\mu'_N} [A^{[m_{N-1}]}_{N-1}]_{\mu \mu_{N-1}} [A^{[m_{N}]}_N]_{\mu_{N-1}\mu_N}}} = 0,\;\;\forall\;\mu_N,\mu'_N, 
\label{eq:CNreqdcondtensor}
\end{equation}
where we have used the orthonormalization conditions of Eq.~(\ref{eq:CNdefnnorm}) for the vectors $\{\oket{C^{(N-2)}_{\mu_{N-2}}}\}$.  
The condition of Eq.~(\ref{eq:CNreqdcondtensor}) is shown diagrammatically in Fig.~\ref{fig:MPScomm}(c).
Since the tensor $A_{N-1}$ is known from the previous step of the recursion, we can use Eq.~(\ref{eq:CNreqdcondtensor}) to solve for the tensor $A_N$. 
To do so, we note that Eq.~(\ref{eq:CNreqdcondtensor}) is equivalent to
\begin{equation}
    \sumal{\mu'_{N-1}, m'_N}{}{\sumal{\mu_{N-1}, m_N}{}{[A^{[m'_N]}_N]^\ast_{\mu'_{N-1}\mu'_N} M^{[m'_N, m_N]}_{\mu'_{N-1}, \mu_{N-1}} [A^{[m_N]}_N]_{\mu_{N-1}\mu_N}}} = 0,\;\;\forall \mu_N, \mu'_N, 
\label{eq:eigprob}
\end{equation}
where we have defined 
\begin{equation}
    M^{[m'_N, m_N]}_{\mu'_{N-1}, \mu_{N-1}} \defn  \sumal{\mu}{}{\sumal{m_{N-1}, m'_{N-1}}{}{\obra{m'_{N-1} m'_N} \hmP_{N-1, N} \oket{m_{N-1} m_N} [A^{[m'_{N-1}]}_{N-1}]^\ast_{\mu\mu'_{N-1}} [A^{[m_{N-1}]}_{N-1}]_{\mu \mu_{N-1}}}},
\label{eq:effectivematrix}
\end{equation}
which is pictorially shown in Fig.~\ref{fig:MPScomm}(d).
Note that we can introduce composite indices $(m'_N, \mu'_{N-1})$ and $(m_N, \mu_{N-1})$ in Eqs.~(\ref{eq:eigprob}) and (\ref{eq:effectivematrix}) and view $A_N$ and $M$ as matrices of dimensions $(d^2_{\loc} \chi_{N-1}) \times \chi_N$ and $(d^2_{\loc}\chi_{N-1}) \times (d^2_{\loc} \chi_{N-1})$ respectively.
Further, it is easy to check using Eq.~(\ref{eq:effectivematrix}) that $M^{[m'_N, m_N]}_{\mu'_{N-1}, \mu_{N-1}} = \big(M^{[m_N, m'_N]}_{\mu_{N-1}, \mu'_{N-1}} \big)^*$, hence $M$ with these composite indices is a Hermitian matrix. 
Moreover, since $\hmP_{N-1, N} = \hmL^\dagger_{N-1, N} \hmL_{N-1,N}$ (explicitly given in our setting but also true for any Hermitian positive semi-definite $\hmP_{N-1, N}$), it is easy to see that the matrix $M$ can be expressed as $M = G^\dagger G$ for an appropriately defined matrix $G$ of shape $\chi_{N-2} d^4_{\loc} \times d^2_{\loc}\chi_{N-1}$; this is also evident from Fig.~\ref{fig:MPScomm}(d). 
Hence $M$ is a Hermitian positive semi-definite matrix, and the tensor $A_N$ in Eq.~(\ref{eq:eigprob}) is the nullspace of $M$, since $A_N$ with composite indices is a $(d^2_{\loc} \chi_{N-1}) \times \chi_N$ matrix where the columns are vectors that make up the nullspace of the matrix $M$. 
The tensor $A_N$ can thus be constructed by diagonalizing a $(d^2_{\loc} \chi_{N-1})$-dimensional matrix. 
The dimension of the kernel of the matrix $M$ determines the $\chi_N$, which is also the dimension of the commutant $\mC_N$. 
Note that in this method, the tensor $A_N$ can be obtained with only the knowledge of the tensor $A_{N-1}$ and the strictly local term $\hH_{N-1, N}$.
[Note that the assumed orthonormalization of the prior set $\{\oket{C^{(N-2)}_{\mu_{N-2}}}\}$ is automatically ensured in the recursive construction starting from Eq.~(\ref{eq:twositeschmidt}) and subsequent diagonalizations of Hermitian matrices $M$ at each step.] 
The full commutant $\mC_N$ can be constructed from the tensors $\{A_j\}_{j = 1}^N$ using Eq.~(\ref{eq:CNMPSform}), although storing it explicitly is typically memory intensive.
\subsection{Bond algebras generated by strictly local \texorpdfstring{$r$}{}-site terms with OBC}\label{subsec:nbondalgebracomm}
The method presented in App.~\ref{subsec:2bondalgebracomm} can be generalized straightforwardly to bond algebras generated by $r$-site terms with OBC.
In this case, the generators of the algebra $\{\hH_{\alpha}\}$ and the superoperators $\{\hmP_{\hH_\alpha}\}$ can be denoted by $\{\hH_{[j,j+r-1]}\}_{j = 1}^{L-r+1}$ and $\{\hmP_{[j,j+r-1]}\}_{j = 1}^{L-r+1}$.
The tensors $\{A_j\}_{j = 1}^r$ can be obtained by solving for the kernel of $\hmP_{[1, r]}$ and performing successive Schmidt decompositions similar to the $r = 2$ case shown in Eq.~(\ref{eq:twositeschmidt}). 
The remaining tensors can be obtained recursively by imposing the conditions of Eq.~(\ref{eq:CNreqdcondition}) for $\obra{C^{(N)}_{\alpha'}}\hmP_{[N - r +1, N]}\oket{C^{(N)}_\alpha} = 0$.
In particular, expressing $\oket{C^{(N)}_{\mu_N}}$ in terms of $\{\oket{C^{(N-r)}_{\mu_{N-r}}}\}$ and the tensors $\{A_j\}_{j = N - r + 1}^N$ [similar to Eq.~(\ref{eq:CNtwice}) for $r = 2$], this condition can be written in terms of the matrix elements of $\hmP_{[N-r+1,  N]}$ and the tensors $\{A_j\}_{j = N - r + 1}^N$ [similar to Eq.~(\ref{eq:CNreqdcondtensor}) for $r = 2$].
Going through steps similar to Eqs.~(\ref{eq:CNreqdcondtensor}) to (\ref{eq:effectivematrix}), we can express $A_N$ in terms of the kernel of a Hermitian positive semi-definite ($d^2_{\loc} \chi_{N-1}$)-dimensional matrix $M$, and the dimension of this kernel is given by $\chi_N$, the dimension of the commutant $\mC_N$. 
Hence the tensor $A_N$ can be determined with the knowledge of the previous $(r-1)$ tensors $\{A_j\}_{j = N - r + 1}^{N-1}$ and the term $\hmP_{[N - r+ 1, N]}$, and the full commutant $\mC_N$ can be constructed from the tensors $\{A_j\}_{j = 1}^N$. 
This method can also be extended to cases where $\{\hmP_{\hH_\alpha}\}$ consist of multiple types of strictly local terms of various ranges.
We can then ``absorb" the smaller range terms into the longer range ones while ensuring that all terms in $\{\hmP_{\hH_\alpha}\}$ have been included, and apply the same procedure.
For example, given two terms of $\hmP^{(1)}_{[j, j+ r_1 - 1]}$ and $\hmP^{(2)}_{[j, j+ r_2 - 1]}$ of ranges $r_1$ and $r_2$ where $r_1 \geq r_2$, we can replace these terms by a new term of range $r_1$, e.g., $\hmP^{(1,2)}_{[j, j+ r_1 - 1]} \defn \hmP^{(1)}_{[j, j+ r_1 - 1]} + \hmP^{(2)}_{[j, j+ r_2 - 1]}$.  
The kernel of $\hmP^{(1,2)}_{[j,j+r_1 - 1]}$ is guaranteed to be the common kernel of $\hmP^{(1)}_{[j,j+r_1 - 1]}$ and $\hmP^{(2)}_{[j, j + r_2 - 1]}$ since they are positive semi-definite operators. 
This procedure is also useful in cases where the ranges of $\{\hmP_{\hH_\alpha}\}$ vary throughout the system, e.g., when the generators include additional shorter range terms on the boundaries of the system.
\begin{figure*}
\centering
\includegraphics[scale=0.45]{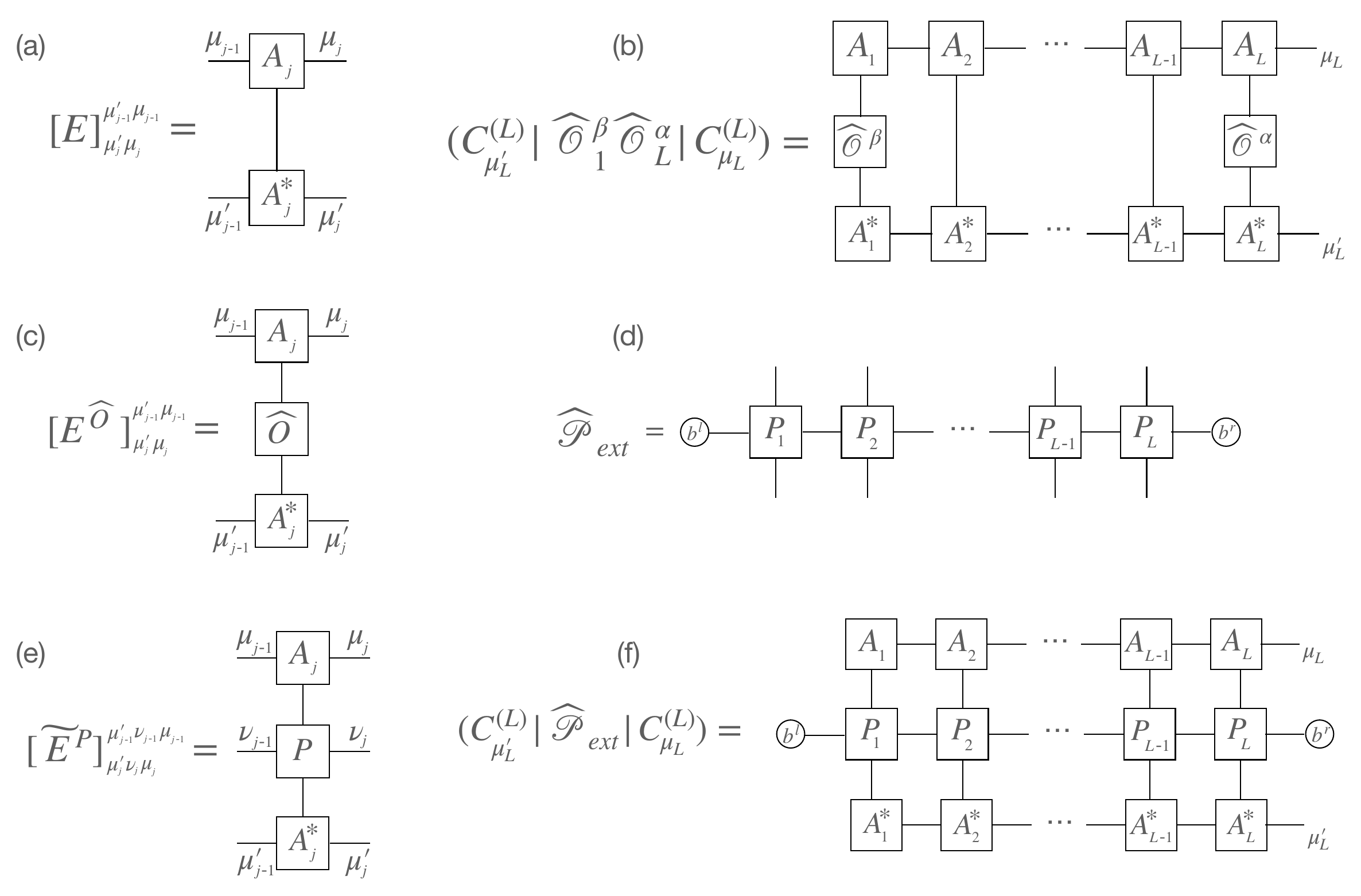}
\caption{
Diagrammatic representation of various tensors required to efficiently construct commutants for bond algebras in one dimension with periodic boundary conditions and for local algebras in one dimension.
}
\label{fig:MPScomm2}
\end{figure*}
\subsection{Bond algebras with strictly local generators and  PBC}\label{subsec:PBCbondalgebra}
We now consider the case where the bond algebra generators $\{\hH_\alpha\}$ consist of strictly local terms with PBC, e.g., terms such as $\hH_{L, 1}$ straddling the ``boundary" in addition to $\{\hH_{j,j+1}\}_{j = 1}^{L-1}$.
In many cases, the addition of this straddling term does not give rise to a new bond algebra, e.g., in the free-fermion and Hubbard algebras discussed in \cite{moudgalya2022from}, although this might not be evident a priori. 
In the following we refer to the commutant of the bond algebra $\lgen \{\hH_{j,j+1}\}_{j= 1}^{L-1} \rgen$ as the OBC commutant and the commutant of the bond algebra $\lgen \{\hH_{j,j+1}\}_{j= 1}^{L-1}, \hH_{L, 1} \rgen$ as the PBC commutant. 
For simplicity, we restrict ourselves to bond algebras generated by nearest neighbor terms, the generalization to other types of terms is straightforward. 
For a system of size $L$ we first compute the OBC commutant $\mC_L$ using the methods in App.~\ref{subsec:2bondalgebracomm}, and its basis vectors $\{\oket{C^{(L)}_{\mu_L}}\}$ which have an MPS form of Eq.~(\ref{eq:CNMPSform}).
The PBC commutant then consists of linear combinations of $\{\oket{C^{(L)}_{\mu_L}}\}$ that are annihilated by the straddling superoperators $\hmP_{\strad}$, e.g., $\hmP_{L, 1}$.
These linear combinations are in the kernel of the $\chi_L$-dimensional Hermitian and positive semi-definite matrix $\hmT_{\strad}$, the restriction of $\hmP_{\strad}$ to $\mC_L$, whose elements are given by
\begin{equation}
    (\hmT_{\strad})_{\mu'_L\mu_L} \defn \obra{C^{(L)}_{\mu_L'}} \hmP_{L, 1} \oket{C^{(L)}_{\mu_L}}. 
\label{eq:Tmatrix}
\end{equation}
While it is memory intensive to compute the vectors $\{\oket{C^{(L)}_{\mu_L}}\}$, the matrix elements can nevertheless be computed efficiently using the MPS form for $\{\oket{C^{(L)}_{\mu_L}}\}$ using the ``transfer matrices" $\{E_j\}$ and $\{E^{\hO}_j\}$ corresponding to the tensors $\{A_j\}$, defined as 
\begin{equation}
    [E_j]^{\mu'_{j-1} \mu_{j-1}}_{\mu'_j \mu_j} \defn \sumal{m_j = 1}{d^2_{\loc}}{[A^{[m_j]}_j]^\ast_{\mu_{j-1}' \mu_j'} [A^{[m_j]}_j]_{\mu_{j-1} \mu_j}},\;\;\;[E^{\hO}_j]^{\mu'_{j-1} \mu_{j-1}}_{\mu'_j \mu_{j}} \defn \sumal{m_j, m'_j = 1}{d^2_{\loc}}{
    [A^{[m'_j]}_j]^\ast_{\mu_{j-1}^{'} \mu_j^{'}} \obra{m_j'}\hO\oket{m_j}
    [A^{[m_j]}_j]_{\mu_{j-1} \mu_j}},
\label{eq:Ejmatrix}
\end{equation}
where $\hO$ is an operator acting on the site $j$. 
Note that these transfer matrices can be viewed as $\chi^2_{j-1} \times \chi^2_j$ matrices by introducing composite indices $(\mu_{j-1}', \mu_{j-1})$ and $(\mu'_j, \mu_j)$, and they are shown diagrammatically in Figs.~\ref{fig:MPScomm2}(a,c). 
Decomposing the straddling operator as $\hmP_{L, 1} = \sum_{\alpha,\beta}{\hmO^\alpha_L \hmO^\beta_1}$, it is easy to see using Eqs.~(\ref{eq:CNMPSform}) and (\ref{eq:Ejmatrix}) that the matrix elements of $\hmT_{\strad}$ of Eq.~(\ref{eq:Tmatrix}) can be expressed as
\begin{equation}
    (\hmT_{\strad})_{\mu'_L, \mu_L} = \sumal{\alpha,\beta}{}{\obra{C^{(L)}_{\mu'_L}} \hmO^{\beta}_1 \hmO^{\alpha}_L \oket{C^{(L)}_{\mu_L}}} = \sum_{\alpha,\beta}[E^{\hmO^\beta}_1 E_2 E_3 \cdots E_{L-1} E^{\hmO^\alpha}_L]_{\mu'_L, \mu_L}^{\mu'_0,\mu_0}.
\label{eq:Tmatrix_via_Es}
\end{equation}
where the equality of the matrix element on the L.H.S. to the transfer matrix expression on the R.H.S. is evident from Fig.~\ref{fig:MPScomm2}(b) (we have indicated $\mu_0', \mu_0$ for clarity, but these are fixed $\mu'_0 = \mu_0 = 1$ and can be dropped, the convention being $[A_1^{[m_1]}]_{\mu_0,\mu_1} = [A_1^{[m_1]}]_{1,\mu_1} \equiv [A_1^{[m_1]}]_{\mu_1}$).
Note that if the operators $\mO^\alpha_L$ and $\mO^\beta_L$ are fermionic, i.e., are odd under fermion parity, one would need to introduce a Jordan-Wigner string that runs throughout the system that enters into the transfer matrix expression in Eq.~(\ref{eq:Tmatrix_via_Es}).
The PBC commutant can then be expressed in terms of the vectors $\{\oket{C^{(L)}_{\mu_{L}}}\}$ by diagonalizing the $\chi_L$-dimensional matrix $\hmT_{\strad}$.
This method can also be applied to bond algebras with multiple terms straddling the boundary, e.g., in the case of bond algebras generated by $n$-site terms for $n > 2$. 
The $\hmT_{\strad}$ matrices similar to Eq.~(\ref{eq:Tmatrix}) can be constructed separately for each type of straddling term, and the full PBC commutant is the kernel of the sum of the matrices, which is guaranteed to be the common kernel of the individual matrices since they are positive semi-definite.
\subsection{Local algebras with some extensive local generators}\label{subsec:localalgebra}
Finally, we can also apply similar ideas to local algebras where the list of generators includes an extensive local term, say $\hH_{\ext}$. 
The idea is again to first compute the OBC commutant $\mC_L$ and the MPS form of its basis vectors $\{\oket{C^{(L)}_{\mu_L}}\}$.
We then compute the matrix elements of the $\hmP_{\ext}$ superoperator of Eq.~(\ref{eq:psdLiouv}) corresponding to the extensive local operator $\hH_{\ext}$ between the vectors that span the OBC commutant:
\begin{equation}
    (\hmT_{\ext})_{\mu'_L\mu_L} \defn \obra{C^{(L)}_{\mu'_L}}\hmP_{\ext}\oket{C^{(L)}_{\mu_L}}.
\label{eq:Textmatrix}
\end{equation}
To compute this overlap, it is convenient to represent $\hmP_{\ext}$, which is translation-invariant if $\hH_{\ext}$ is translation-invariant, as a Matrix Product Operator (MPO)~\cite{pirvu2010matrix}, i.e.,
\begin{equation}
\hmP_{\ext} = \sumal{\{m'_j\}, \{m_j\}}{}{[{b^l}^T P^{[m'_1 m_1]} P^{[m'_2 m_2]} \dots P^{[m'_L m_L]} b^r]}\oket{m'_1 \cdots m'_L}\obra{m_1\cdots m_L},
\label{eq:PMPOform}
\end{equation}
where the $\{P^{[m'_k ,m_k]}\}$ are $\Upsilon$-dimensional matrices, and $b^l$, $b^r$ are $\Upsilon$-dimensional vectors whose elements can be chosen to be some fixed numbers. 
We pictorially show the MPO form of Eq.~(\ref{eq:PMPOform}) in Fig.~\ref{fig:MPScomm2}(d).
Note that since $\hmP_{\ext} \defn \hmL^\dagger_{\ext} \hmL_{\ext}$, the MPO matrices $\{P^{[m'_k,m_k]}\}$ can be constructed from the MPO matrices for $\hmL_{\ext}$.
This can in turn be constructed directly from the expression of the terms in $\hH_{\ext}$.
For example, if $\hH_{\ext} = \sum_{j}{\hh_{[j,j+r]}}$, where $\{\hh_{[j,j+r]}\}$ are some range-$r$ strictly local operators, $\hmL_{\ext}$ is given by
\begin{equation}
\hmL_{\ext} = \hH_{\ext} \otimes \mathds{1} - \mathds{1} \otimes \hH_{\ext}^T = \sum_j{(\hh_{[j, j+r]} \otimes \mathds{1} - \mathds{1} \otimes \hh_{[j, j+r]})}.
\label{eq:Lextexample}
\end{equation}
This shows that $\hmL_{\ext}$ is a sum of range-$r$ strictly local superoperators, and its MPO can be constructed using several standard techniques known in the literature~\cite{crosswhite2008fsa, motruk2016density, moudgalya2018entanglement}. 
For example, if $\hH_{\ext} = S^z_{\tot} = \sum_j{S^z_j}$, which is the extensive local operator we are interested in several examples, it is easy to show that $\hmL_{\ext}$ is an MPO of bond dimension $2$, i.e.,
\begin{equation}
\hmL_{\ext} = \sumal{\{m'_j\}, \{m_j\}}{}{[{b^l_Q}^T Q^{[m'_1 m_1]} Q^{[m'_2 m_2]} \dots Q^{[m'_L m_L]} b^r_Q]}\oket{m'_1 \cdots m'_L}\obra{m_1\cdots m_L},
\label{eq:LMPOform}
\end{equation}
where $Q$ is the MPO tensor which can be viewed as a 2-dimensional matrix with elements as single-site superoperators, and $b^l_Q$, $b^r_Q$ are 2-dimensional vectors; their expressions read
\begin{equation}
Q = \begin{pmatrix} \mathds{1} \otimes \mathds{1} & S^z \otimes \mathds{1} - \mathds{1} \otimes S^z \\
0 & \mathds{1} \otimes \mathds{1} \end{pmatrix},\;\;b^l_Q = \begin{pmatrix}
1 \\ 0 \end{pmatrix}, \;\;b^r_Q = \begin{pmatrix} 0 \\ 1 \end{pmatrix}.
\label{eq:LMPOtensors}
\end{equation}
Coming back to the computation of Eq.~(\ref{eq:Textmatrix}), we can define generalized transfer matrices as
\begin{equation}
    [\tE^{P}_j]^{\mu'_{j-1}\nu_{j-1}\mu_{j-1}}_{\mu'_{j}\nu_{j}\mu_{j} } \defn \sumal{m'_j, m_j = 1}{d^2_{\loc}}{[A^{[m'_j]}_j]^\ast_{\mu_{j-1}' \mu_j'}[P^{[m'_j, m_j]}]_{\nu_{j-1}, \nu_j} [A^{[m_j]}_j]_{\mu_{j-1} \mu_j} },
\label{eq:Etildjmatrix}
\end{equation}
which is shown in Fig.~\ref{fig:MPScomm2}(e) and can be viewed as a $(\chi^2_{j-1} \Upsilon \times \chi^2_j \Upsilon)$ matrix.
We can then express the matrix elements of $\hmT_{\ext}$ of Eq.~(\ref{eq:Textmatrix}) as
\begin{equation}
    (\hmT_{\ext})_{\mu'_L, \mu_L} = \sumal{\nu_0, \nu_L}{}{(b^l)_{\nu_0} [\tE_1 \tE_2 \cdots \tE_{L-1} \tE_L]^{\mu'_0, \nu_0, \mu_0}_{\mu'_L, \nu_L, \mu_L} (b^r)_{\nu_L}},
\label{eq:Texttransfer}
\end{equation}
where the boundary vectors $b^l$ and $b^r$ only carry the auxiliary indices of the MPO, as shown in Fig.~\ref{fig:MPScomm2}(f) [and $\mu'_0 = \mu_0 = 1$ are fixed and can be dropped as explained after Eq.~(\ref{eq:Tmatrix_via_Es})].
The commutant of this local algebra is then given by the kernel of the Hermitian positive semi-definite matrix $\hmT_{\ext}$.
\end{document}